\pdfoutput=1

\documentclass[prb, onecolumn, showpacs]{revtex4}

\usepackage{enumitem}
\usepackage{graphicx}
\usepackage{bbm}
\usepackage{amssymb,amsmath}
\usepackage{appendix}
\usepackage{epsfig}
\usepackage{color}
\usepackage{helvet}
\def\be{\begin{equation}}
\def\ee{\end{equation}}
\def\bea{\begin{eqnarray}}
\def\eea{\end{eqnarray}}

\begin{document}
\title{Signal-to-pump back-action and self-oscillation in Double-Pump Josephson Parametric Amplifier}
\author{Archana Kamal}
\email{archana.kamal@yale.edu}
\author{Adam Marblestone}
\author{Michel Devoret}
\affiliation{Department of Applied Physics and Physics, Yale University, New Haven, Connecticut 06520-8284, USA}
\begin{abstract}
We present the theory of a Josephson parametric amplifier employing two pump sources. Our calculations are based on Input-Output Theory, and can easily be generalized to any coupled system involving parametric interactions. We analyze the operation of the device, taking into account the feedback introduced by the reaction of the signal and noise on the pump power, and in this framework, compute the response functions of interest - signal and idler gains, internal gain of the amplifier, and self-oscillation signal amplitude. To account for this back-action between signal and pump, we adopt a mean-field approach and self-consistently explore the boundary between amplification and self-oscillation. The coincidence of bifurcation and self-oscillation thresholds reveals that the origin of coherent emission of the amplifier lies in the multi-wave mixing of the noise components. Incorporation of the back-action leads the system to exhibit hysteresis, dependent on parameters like temperature and detuning from resonance. Our analysis also shows that the resonance condition itself changes in the presence of back-action and this can be understood in terms of the change in plasma frequency of the junction. The potential of the double pump amplifier for quantum-limited measurements and as a squeezer is also discussed.
\end{abstract}
\maketitle
%
%
%ccccccccccccccccccccccccccccccc
\section{Introduction}
%ccccccccccccccccccccccccccccccc
%
%
Recent advances in nanoelectromechanical systems (NEMS) \cite{NEMS1, NEMS2, NEMS3} and optical communication technology \cite{OPT1, OPT2} have renewed interest in the field of quantum-limited amplification and encouraged scientists working in these areas to design systems which can operate at the quantum limit. The seminal work of Caves \cite{PhysRevD.26.1817} set the minimum noise added by a phase-preserving amplifier at half a photon referred to the input channel. The same work showed, however, that phase-sensitive amplifiers are not submitted to this limitation and can amplify one of the quadratures noiselessly at the expense of deamplification of the conjugate quadrature. Parametric amplifiers where a strong pump wave at frequency $\omega_{p}$ causes simultaneous generation of signal and idler photons at frequencies $\omega_{S}$ and $\omega_{I}$ respectively, are particularly promising candidates to reach the quantum limit. These may employ either a second-order nonlinear interaction with $\omega_{p} = \omega_{S} + \omega_{I}$ or a third-order interaction with $2 \omega_{p} = \omega_{S} + \omega_{I}$. Various schemes based on quantum optical parametric amplifiers \cite{1071660,Caves:87, opticsparamp, 1016354, 1207224} operating at sub-quantum noise limits have been tested successfully.
\par
Low noise amplifiers in the RF and microwave domains have seen a growing demand from applications in astro-particle physics  \cite{1201803, RevModPhys.75.777} and quantum information processing \cite{RSL1} involving microwave investigation of solid state qubits. Parametric amplifiers based on Josephson tunnel junctions (JTJ) are a front-runner for such applications as JTJs are the only known non-dissipative and nonlinear circuit elements, operating at radio frequencies and arbitrarily low temperatures. Experimental efforts \cite{PhysRevLett.67.1411, PhysRevLett.93.207002, metcalfe:174516, tholen:253509, castellanos-beltran:083509, yamamoto:042510} have engineered the nonlinearity of JTJ and utilized it to perform challenging measurements like non-demolition readout of superconducting qubits \cite{boulant:014525}.
\par
\begin{figure}[h!]
\includegraphics[width=0.4\textwidth]{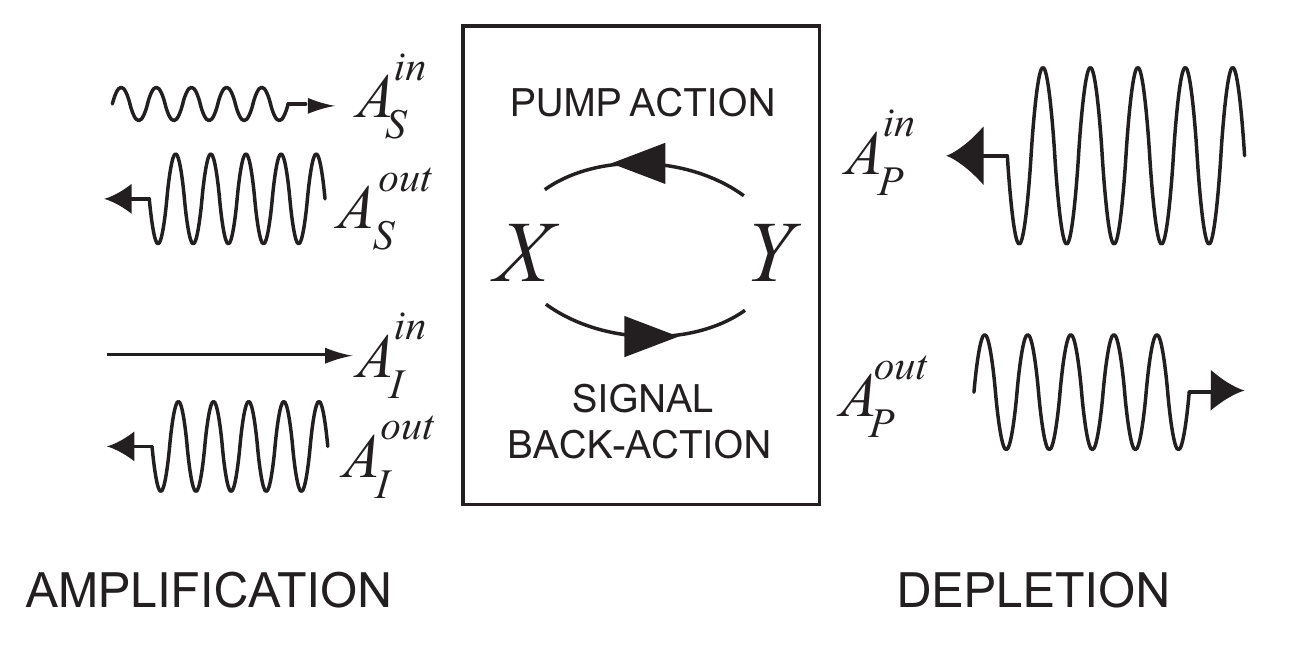}\\
\caption{Scattering representation of a parametric amplifier with its signal, idler and pump ports as well as its internal coordinates $X$ and $Y$. The bidirectional coupling between the internal coordinate at signal frequency ($X$) and that oscillating at the pump frequency ($Y$) is shown using the curved arrows. This coupling leads, in one direction (left arrow) to amplification of the outgoing wave signal while in the other direction (right arrow) to depletion of the outgoing pump wave. It may be noted that this process also leads to a deterministic signal of finite amplitude at the idler port as the pump also populates the outgoing idler wave along with the signal, even if there is no idler input.}
\label{BAdefinition}
\end{figure}

A typical parametric amplification process is depicted in Fig. \ref{BAdefinition}. The amplifier can be visualized as a black-box where an incoming signal ($A^{\textrm{in}}_{S}$) is reflected into a larger signal ($A^{\textrm{out}}_{S}$) due to energy transfer from the pump drive to the signal. Similarly the idler port, empty of incoming waves, gives rise to a spontaneously emitted signal. The interaction between the signal and pump waves can be understood as a coupling between the internal coordinates of the amplifier representing the signal/idler ($X$) and pump ($Y$) modes respectively. Under the stiff pump approximation, the internal coordinate $Y$ is considered to be completely enslaved to the pump amplitude $A^{\textrm{in}}_{P}$. As the noise at the input grows, one eventually enters the soft pump regime where the pump coordinate $Y$ has its dynamics partly determined by $X$. This interaction between the pump and the signal leads to a reciprocal effect of the signal on the pump - referred to as `back-action' in the following text. As a result of this back-action, the outgoing pump wave ($A^{\textrm{out}}_{P}$) suffers depletion, which is the counterpart of amplification of the signal and idler waves. We have analyzed the soft pump regime and the effect of two-way coupling between signal/idler and pump channels in a detailed manner for a parametric amplifier using a single tunnel junction by using a model employing two pump sources. Interestingly, the use of dual-pump scheme simplifies the dynamics of this nonlinear coupled system significantly, a reasonable trade-off at the cost of a second generator. The absence of this crucial simplification makes a similar soft pump analysis for conventional single pump devices unwieldy.\footnote{Our analysis can serve as a model for treating more complex circuits involving multiple junctions driven by a single pump.}
\par
The quantum optics community is well-acquainted with the use of dual pumps in parametric amplification \cite{Yariv, Boggio:01, 1506859}. The Josephson dual-pump amplifier (DPA), besides providing an efficient theoretical tool for studying various aspects of the dynamics of the amplification process, also offers practical advantages like greater bandwidth \cite{twopumpopt}, tunable band center and additional degrees of design freedom. In the following sections, we provide a theoretical study of this novel Josephson ``paramp" and also explore the boundary between amplification and delicate effects like self-oscillation.
\par
\begin{figure}[h!]
\includegraphics[width=0.7\textwidth]{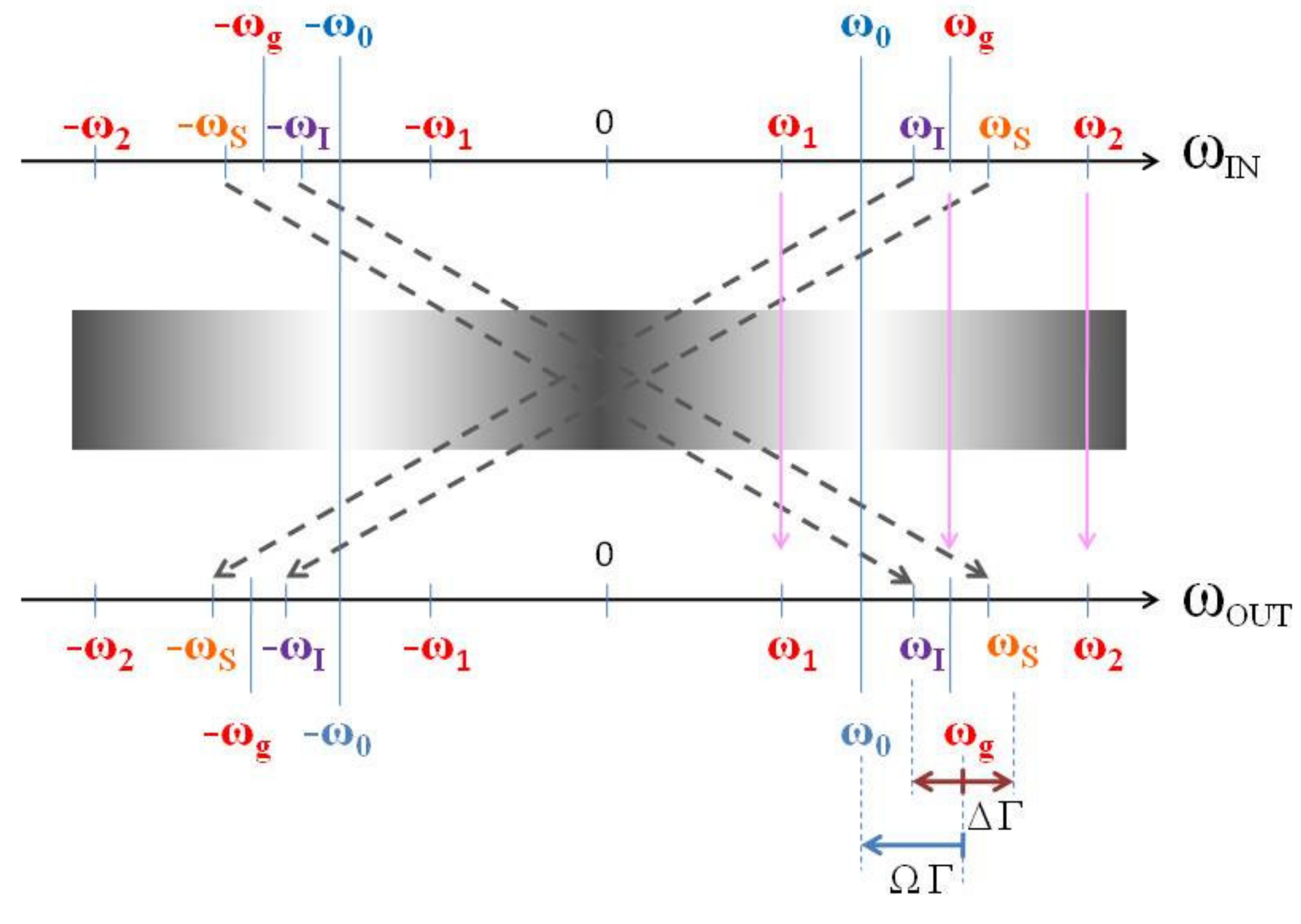}\\
\begin{flushleft}
\caption{Characteristic frequencies of a double-pump amplifier (DPA). The upper/lower axes represents the frequencies at the input/output port. Diagonal dashed arrows represent the parametric coupling between various frequency components. The shade of the gradient in the middle band is a schematic representation of the oscillator bandwidth $2 \Gamma$. The resonant frequency of the oscillator is $\omega_{0}$. The pump frequencies are denoted by $\omega_{1}$ and $\omega_{2}$ respectively and their average $\omega_{g}$ defines the band center. The reduced detuning of $\omega_{0}$ from $\omega_{g}$ is $\Omega = (\omega_{0} - \omega_{g})/\Gamma$. The generic frequency of the signal and that of the corresponding idler are denoted by $\omega_{S}$ and $\omega_{I} = 2\omega_{g} -\omega_{S}$, respectively. The reduced detuning of the signal (idler) from $\omega_{g}$ is $\Delta = (\omega_{S} - \omega_{g})/\Gamma$ ($-\Delta$ for the idler). The reduced detunings $\Omega$ and $\Delta$ are not indicated to scale and have been exaggerated for clarity.}\label{freqspec}
\end{flushleft}
\end{figure}

In a regular degenerate paramp the signal is located within a cavity linewidth of the pump frequency which is not favorable for the observation of fragile effects like self-oscillation. With a double pump scheme, the pump frequencies lie towards outer edges of the band of amplification, leading to better separation of the pump and the signal frequencies (Fig. \ref{freqspec}). This provides a zero background for detection of a small signal like self-oscillation which would have been otherwise challenging due to the large background of the pump. The use of dual pumping further helps in the realization of a symmetric phase boundary for bifurcation. This should be contrasted with a single pump device like the Josephson Bifurcation Amplifier (JBA) \cite{PhysRevLett.93.207002}, where the phase boundary is canted due to a first-order transition between dynamical states of the junction near threshold (cf. Appendix \ref{JBA_app}). As a consequence, unlike the JBA, there is no inherent frustration between the thresholds for self-oscillation and parametric gain in the double-pump amplifier discussed here. It is this symmetry, unique to the DPA, that proves extremely valuable for the simplification of equations governing the dynamics of the system near bifurcation. We deal with this issue in greater detail in section \ref{sec_st_stiff}.
\par
This article is organized as follows: we first derive the basic equations governing the dynamics of the DPA in section \ref{model} showing the presence of various couplings existing in the system. Then in section \ref{sec_stiff}, we calculate the various response functions of the system for a stiff pump. In particular, we study the response of the system to noise and demonstrate the proficiency of the amplifier for achieving the quantum limited noise temperature in \ref{sec_pert_stiff}. This is followed by a derivation of the steady state response of the system (section \ref{sec_st_stiff}) where we show that the system exhibits self-oscillation beyond the bifurcation threshold. Also, we describe the use of the DPA as an efficient squeezer in section \ref{sec_squ_stiff}. In the second part of the paper, we present a complete solution of the problem, accounting for the depletion of the pump due to back-action. We rederive the signal and self-oscillation amplitudes in the presence of back-action in section \ref{sec_ba}, using a mean-field approach. In section \ref{corr_calc}, we arrive at explicit expressions quantifying the magnitude of back-action in terms of system variables. Using the derived value of back-action we plot the corrected values of various response functions of the system. Finally we discuss the implications of our results and offer future perspectives in section \ref{summary}.
\par
%
%
%ccccccccccccccccccccccccccccccc
\section{Basic Model of the DPA}
\label{model}
\begin{figure}[h!]
\includegraphics[width=0.7\textwidth]{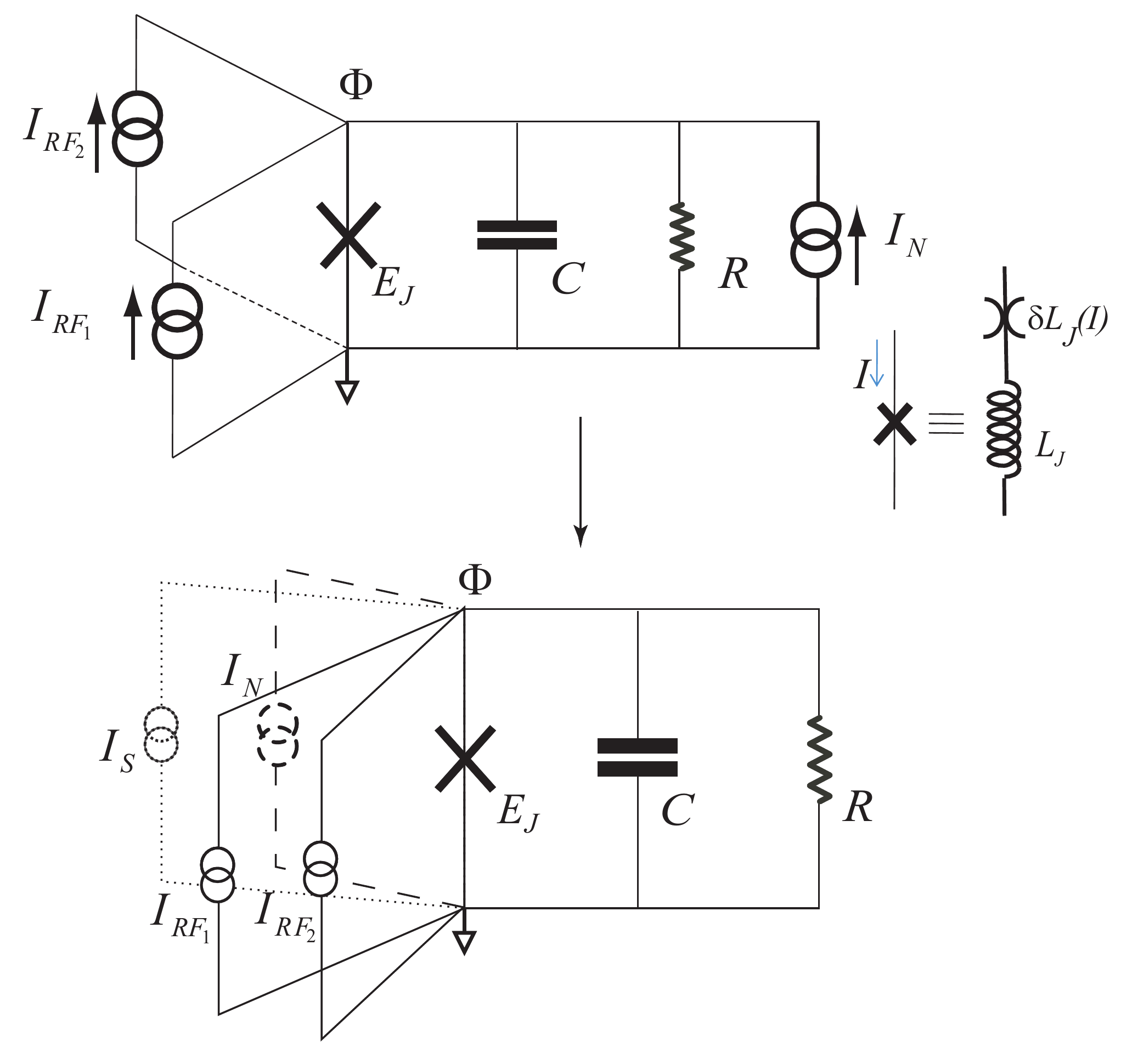}\\
\caption{Top panel: Josephson nonlinear oscillator pumped with two current sources $I_{RF_{1}}$ and $I_{RF_{2}}$. The Josephson element, denoted by a cross symbol, is used for its nonlinear inductance ($\delta L_{J} = (L_{J}/6)(L_{J}^2 I^{2})/ (\hbar/2 e)^{2} + O (I^{2})$) which performs coherent multiwave mixing. Bottom panel represents full schematic of the DPA. The current $I_{S}$ corresponds to the input signal while $I_{N}$ represents the noise current due to the resistance $R$, modeling the internal resistances of current sources. The symbol $\Phi$ is the node flux associated with the Josephson element.}
\label{circ}
\end{figure}

%ccccccccccccccccccccccccccccccc
%
%
A circuit realization of the DPA is depicted in Fig. \ref{circ}. It consists of a nonlinear LC oscillator pumped by two RF currents denoted by $I_{RF_{1}}(t)$ and $I_{RF_{2}}(t)$. The nonlinear inductance of the circuit is provided by a Josephson tunnel junction which can be modeled as a linear inductance $L_{J} = \frac{\hbar}{2 e I_{0}}$ in series with a current-dependent inductance $\delta L_{J} = \lambda L_{J} \frac{I^2}{I_{0}^2}$. Here we neglect higher order nonlinear terms since, for simplicity, the amplifier is operated in the weakly nonlinear regime. Note that our treatment can be applied to more general oscillating Josephson circuit by simply renormalizing the value of $\lambda$ (for instance in the Cavity Bifurcation Amplifier (CBA)\cite{CBA} or even an array of Josephson junctions inside a cavity \cite{castellanos-beltran:083509}).
\par
Equating the currents at the node and expressing them in terms of the node flux $\Phi$, we get the following equation of motion. It is the equation for a nonlinear, damped driven Duffing oscillator. Note that we have expanded the sine function in JTJ current to third order for obtaining this equation.
\begin{eqnarray}
     \ddot{\Phi} + 2 \Gamma \dot{\Phi} + {\omega_{0}}^2\Phi \left[1 + \lambda {\left(\frac{2 \pi \Phi}{\Phi_{0}}\right)}^2\right] = \frac{1}{C} \left(I_{RF_{1}}\cos(\omega_{1}t)+ I_{RF_{2}}\cos(\omega_{2}t) + I_{N}(t)\right)
     \label{master}
\end{eqnarray}
The source terms on the right are the drive currents - $I_{RF_{1}}$, $I_{RF_{2}}$, contributed by two pumps and the `noise current' $I_{N}$, whose presence is imposed by the fluctuation-dissipation theorem.
The damping constant $\Gamma = 1/2 R C$ and the natural frequency of oscillation of the circuit is $\omega_{0} = 1/\sqrt{L_{J} C}$. Here $\lambda$ denotes the nonlinearity coefficient, whose value is $-1/6$ in the case of Josephson junction (cf. Fig. \ref{circ}). Also, the pump frequencies $\omega_{1}$ and $\omega_{2}$ are separated from $\omega_{0}$ by several linewidths $\Gamma$.
\par
\begin{figure}[h!]
\includegraphics[width=0.7\textwidth]{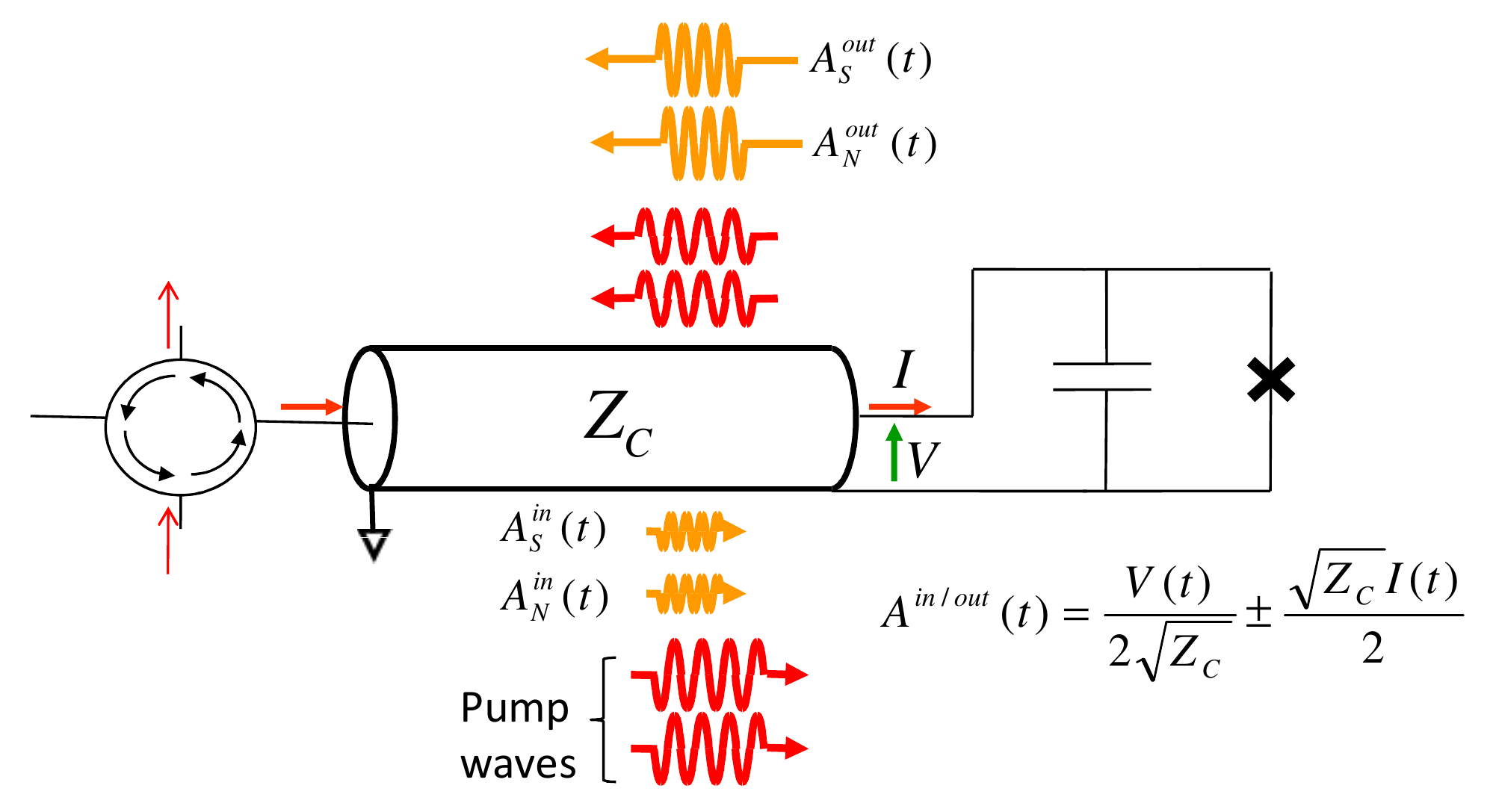}\\
\caption{Traveling wave representation of circuit of Fig. \ref{circ}. The resistance $R$ is replaced with a transmission line of characteristic impedance $Z_{c} = R$ and the voltages and currents are treated as particular linear combinations of incoming and outgoing wave amplitudes.}
\label{IOT}
\end{figure}

To solve the above equation, we will use the formalism of input-output theory (IOT) \cite{IOT} where the electrical voltage and current signals at the level of resistance are treated as fields and are decomposed into incoming and outgoing traveling waves (cf. Fig \ref{IOT}) propagating along a transmission line.
The equations relating the voltage and current at any given position of the line to the wave amplitude ($A$) of the wave propagating along the line are given by (cf. Appendix \ref{IOT_app}):
\begin{eqnarray}
     & & V(t) = V^{\textrm{in}}(t) + V^{\textrm{out}}(t); \; \;\;\;\; I(t) = I^{\textrm{in}}(t) - I^{\textrm{out}}(t)\\
     & & V^{\textrm{in /out}} = \sqrt{Z_{c}} A^{\textrm{in/out}}; \; \; \; \; I^{\textrm{in /out}} = \frac{1}{\sqrt{Z_{c}}} A^{\textrm{in/out}}
\end{eqnarray}
Here the wave amplitude $A$ has dimensions of $(\textrm{Watt})^{1/2}$. We will invoke this technique to solve our final system of equations.
\par
Returning to the main Eq. (\ref{master}), we rewrite it as
\begin{eqnarray}
     \ddot{\varphi} + 2 \Gamma \dot{\varphi} + {\omega_{0}}^2\varphi (1 + \lambda {\varphi}^2)
     - \omega_{0}^2 \frac{I_{RF_{1}}}{I_{0}}- \omega_{0}^2 \frac{I_{RF_{2}}}{I_{0}}= 4 \Gamma v^{\textrm{in}}(t)
     \label{master_dimless}
\end{eqnarray}
where we have introduced a dimensionless variable $\varphi = \frac{2 \pi \Phi}{\Phi_{0}}$. The symbol $v^{\textrm{in}} = \frac{I_{N} R}{\Phi_{0}/2 \pi}$ represents the quantum noise field driving the oscillator in the propagating mode picture. We assume a solution of the form
\begin{eqnarray}
    \varphi = \int_{-\infty}^{\infty} X[\omega] \exp(-\dot{\imath}\omega t) d\omega
    + \{ \Xi \exp(-\dot{\imath}\omega_{g}t) + Y \exp(-\dot{\imath}\omega_{1}t) + Z \exp(-\dot{\imath}\omega_{2}t)+ c.c.\}
\label{sol}
\end{eqnarray}
where $\Xi$ is a phasor corresponding to internal coordinate of amplifier oscillating at frequency $\omega_{g} \left(= \frac{\omega_{1} + \omega_{2}}{2}\right)$. This describes the possible self-oscillation amplitude of the system. The symbols \emph{Y}, \emph{Z} are phasors corresponding to the pumps at frequencies $\omega_{1}$ and $\omega_{2}$ respectively. Note that here we treat the generic signal (at frequency $\omega_{S}$) as a Dirac-delta component $X_{S}[\omega_{S}] = x_{S} \delta (\omega -\omega_{S})+ x_{S}^{*} \delta (\omega + \omega_{S})$ in the noise amplitude $X[\omega]$.
For further use, we introduce the ``noise back-action factor'' $\Pi$ as,
\begin{eqnarray}
    \Pi = \int_{-\infty}^{\infty} \int_{-\infty}^{\infty} d\omega_{a} d\omega_{b} X[\omega_{a}] X[\omega_{b}] \delta(\omega_{a}+\omega_{b}- 2 \omega_{g})
    \label{main5}
\end{eqnarray}
Note that $\Pi$ is dimensionless like $\varphi$ and unlike $X[\omega]$ which has dimensions of time.
\par
Substituting Eq. (\ref{sol}) in (\ref{master_dimless}), using the definition
\begin{eqnarray*}
    f(t) = \int d\omega f[\omega] \exp (-\dot{\imath} \omega t)
\end{eqnarray*}
and performing harmonic balance (for frequencies $\omega_{S}$, $\omega_{I}$, $\omega_{1}$, $\omega_{2}$) leads to the following system of coupled equations for \emph{X}, \emph{Y}, \emph{Z}, $\Pi$ and $\Xi$
\begin{eqnarray}
    & & (-\omega_{S}^2 - 2 \dot{\imath} \Gamma \omega_{S} + \omega_{0}^2) X[\omega_{S}] + \chi X[-\omega_{I}]Y Z
    = 4 \Gamma v^{\textrm{in}}[\omega_{S}]
    \label{main1}\\
    & & (-\omega_{I}^2 - 2 \dot{\imath} \Gamma \omega_{I} + \omega_{0}^2) X[\omega_{I}] + \chi X[-\omega_{S}]Y Z
    = 4 \Gamma v^{\textrm{in}}[\omega_{I}]
    \label{main2}\\
    & & (-\omega_{1}^2 - 2 \dot{\imath} \Gamma \omega_{1} + \omega_{0}^2) Y  + \chi \Pi Z^{*} + \frac{\chi}{2} \Xi^2 Z^{*} = f_{1}
    \label{main3}\\
    & & (-\omega_{2}^2 - 2 \dot{\imath} \Gamma \omega_{2} + \omega_{0}^2)Z + \chi \Pi Y^{*}  + \frac{\chi}{2} \Xi^2 Y^{*} = f_{2}
    \label{main4}\\
    & & (-\omega_{g}^2 - 2 \dot{\imath} \Gamma \omega_{g} + \omega_{0}^2) \Xi + \chi \Xi^{*}[\omega_{g}]Y Z= 0
    \label{main6}
\end{eqnarray}
with $\chi = \lambda \omega_{0}^2, f_{1,2} = \omega_{0}^2 \frac{I_{RF_{1,2}}}{I_{0}}$. Also note that $X[\omega_{I}]$ is the component of input noise at the idler frequency $ \omega_{I} = (\omega_{1} +\omega_{2} - \omega_{S})$.
\par
As evident from the Eq. \ref{main5}, the back-action factor $\Pi$ involves the mixing of signal and idler waves, realized through the nonlinearity of the system. It captures the effect of the reaction due to signal idler coupling on the pumps and eventually leads to pump depletion. Hence, it serves to make the coupling between pumps and signals symmetric in the sense that as the signal grows due to transfer of energy from pump into signal channel, it leads to a reduction of the pump amplitudes (as can be seen from Eqns. \ref{main3} and \ref{main4}) and this depleted pump then acts on the signal. The delta function in frequency in the expression for $\Pi$ is included to enforce harmonic balance (cf. Eqns. (\ref{main3}) and (\ref{main4})).
\par
In deriving the above set of equations we have ignored the $2 \omega_{0}$ components and higher harmonics under the rotating wave approximation (RWA). Also note that we have neglected additional terms of the type $|X|^2 + |Y|^2 + |Z|^2$ in each of the above equations as their effect on dynamics is a simple 'renormalization' of the plasma frequency (it will be seen later in section \ref{sec_st_stiff} that $\omega_{0}$ is actually a function of self-oscillation amplitude $\Xi$ and noise back-action factor $\Pi$.) We have also dropped the terms of the form like $\Pi X^{*}[\omega_{S}]$, denoting explicit coupling between different back-action terms exclusive of the pump, as we will restrict our analysis to the perturbative limit where such terms are much smaller in magnitude. Under the above scheme of approximations, the system of nonlinear coupled Eqns. (\ref{main5}-\ref{main6}) form the basic framework of the problem to be solved.
%
%
%ccccccccccccccccccccccccccccccc
\subsection*{Pump amplitudes: Signature of back-action}
%ccccccccccccccccccccccccccccccc
%
%
We solve Eqns. (\ref{main3}) and (\ref{main4}) simultaneously, in the high-Q limit ($\Gamma \ll \omega_{0}$), to obtain:
\begin{equation}
    \boxed{Y Z = \tilde{f_{1}} \tilde{f_{2}} \left[ - 1 + \gamma \left(\Pi + \frac{1}{2}\Xi^2 \right)\right]}
    \label{main7}
\end{equation}
where $\gamma = \frac{2 \lambda}{(1+\frac{\omega_{1}}{\omega_{0}})(1+\frac{\omega_{2}}{\omega_{0}})},\; \tilde{f_{i}} = \frac{I_{RF_{i}}}{I_{0}}\frac{1}{(1 + \frac{\omega_{i}}{\omega_{0}}) |1 - \frac{\omega_{i}}{\omega_{0}}|}$ are dimensionless constants of the problem. As is evident from the above equation, the correction to the pump amplitudes due to back-action ($\Pi$ and $\Xi^2$) depends mainly on the nonlinearity parameter $\lambda$.
\par
Eq. (\ref{main7}) gives the magnitude of the product $YZ$ instead of individual pump amplitudes. It is convenient to formulate the drive strength in this manner as it is the product of two pump amplitudes which enters the equations of motion of signal and idler (cf. Eqns. (\ref{main1}), (\ref{main2})). Also, it should be noted that as we work in the perturbative limit or weak nonlinearity, we have included terms only up to the linear order in $\lambda$ in deriving the above equation.
%
%cccccccccccccccccccccccccccccccc
\section{Stiff Pump Approximation}
\label{sec_stiff}
%cccccccccccccccccccccccccccccccc
%
%
%ccccccccccccccccccccccccccc
\subsection{Signal and Idler Response}
\label{sec_pert_stiff}
\begin{figure}[h!]
\centering
\includegraphics[width=\textwidth]{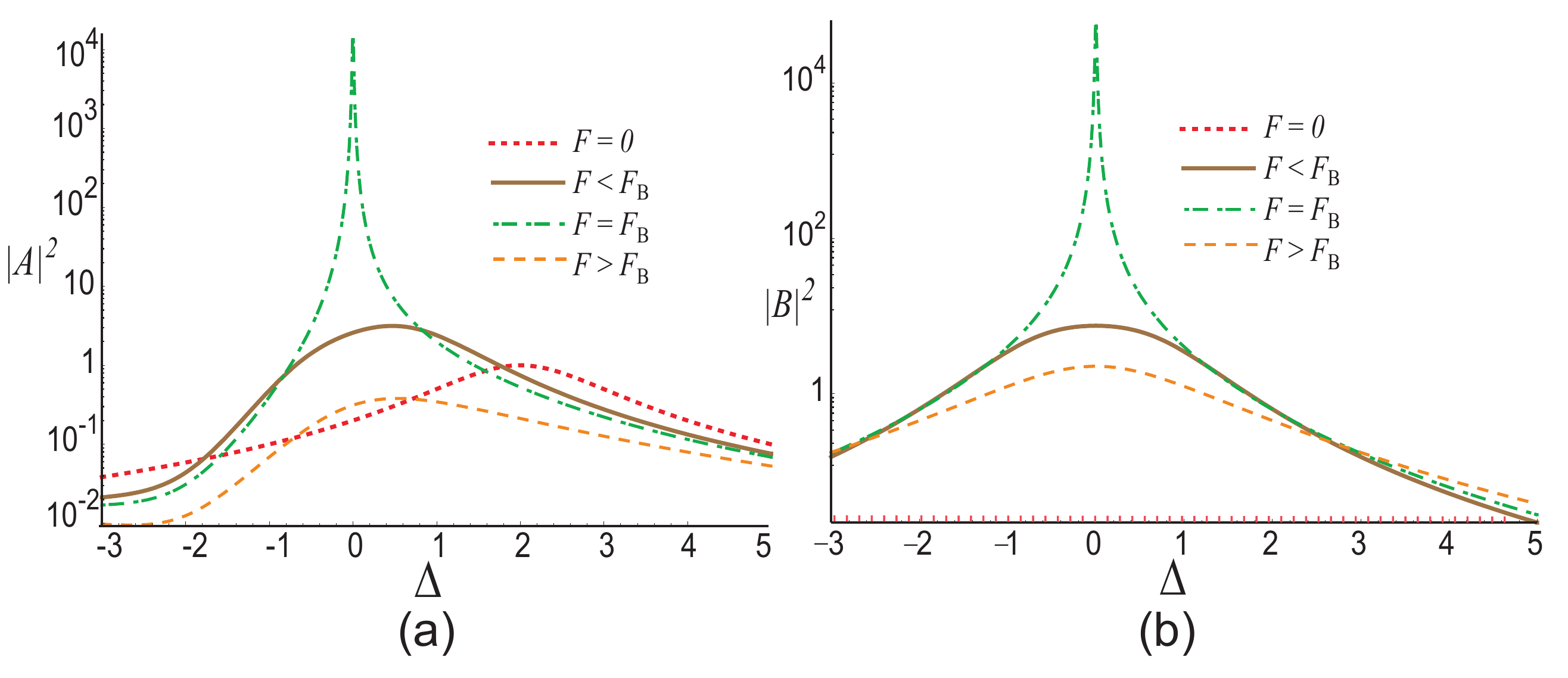}\\
\caption{(a) Signal-to-signal response of the internal coordinate to the incoming signal (internal \emph{cis}-gain of the DPA) and (b) signal-to-idler response of the internal coordinate to the incoming signal (internal \emph{trans}-gain of the DPA) plotted for reduced pump detuning $\Omega = 2$ as a function of reduced signal detuning $\Delta$. The values $\Delta = 2$ and $\Delta = 0$ correspond to $\omega_{S} = \omega_{0}$ and $\omega_{S} = \omega_{g}$, respectively. The two spectra diverge in a similar fashion as the system is driven to the bifurcation threshold power $F_{B}$.}
\label{draw1}
\end{figure}

%ccccccccccccccccccccccccccc
%
%
Under the stiff pump approximation we ignore the effect of the back-action terms $ \gamma\left( \Pi + \frac{1}{2} \Xi^2\right) $ in Eq. (\ref{main7}). Thus, the pump amplitudes are simply given by $YZ = -\tilde{f_{1}}\tilde{f_{2}}$. This approximation will be further validated when we explicitly calculate the self-oscillation signal in subsequent sections. Using this value of drive amplitudes in Eqns. (\ref{main1}) and (\ref{main2}) and requesting $\omega \sim \omega_{0}$ (under RWA), the signal and idler equations yield
\begin{eqnarray}
    (\Omega - \Delta -\dot{\imath})X[\omega_{S}] - F X[-\omega_{I}] = \left(\frac{2}{\omega_{0}}\right)v^{\textrm{in}}[\omega_{S}]\\
    (\Omega + \Delta - \dot{\imath})X[\omega_{I}] - F X[-\omega_{S}] = \left(\frac{2}{\omega_{0}}\right) v^{\textrm{in}}[\omega_{I}]
\end{eqnarray}
$\Delta =\frac{\omega_{S}-\omega_{g}}{\Gamma}$ represents the reduced detuning of the signal from the ghost frequency $\omega_{g}$. In this scheme of measuring the frequencies from the band-center $\omega_{g}$, the idler corresponds to a detuning $-\Delta$ from the ghost frequency. In the following expressions, we switch to the convention where $X[+|\omega|] = X[|\omega|]$ and $X[-|\omega|] = X^{*}[|\omega|]$ in order to avoid confusion between negative frequencies and negative reduced signal detuning.  Also introduced is the reduced pump detuning $\Omega = \frac{\omega_{0}-\omega_{g}}{\Gamma}$, and the \emph{effective pump power}
\begin{equation}
    F \simeq \frac{2 \lambda}{\Gamma/\omega_{0}}\frac{I_{RF1}I_{RF2}}{I_{0}^2}
    \label{DPAdrive}
\end{equation}
(for $\omega_{1} = 3/4 \omega_{0}, \omega_{2} = 5/4 \omega_{0}$).
On solving the two equations simultaneously, we get
\begin{eqnarray}
    X(\Delta)& = & \left(\frac{2}{\omega_{0}}\right) \frac{(\Omega + \Delta +\dot{\imath}) v^{\textrm{in}}(\Delta) +  F v^{*\textrm{in}}(-\Delta)}
    {(\Omega - \Delta - \dot{\imath})(\Omega + \Delta + \dot{\imath}) - F^2}\nonumber\\
    & = & \left(\frac{2}{\omega_{0}}\right)\left[ A(\Delta) \; v^{\textrm{in}}(\Delta) + B(\Delta)\; v^{*\textrm{in}}(-\Delta)\right] \label{cis}\\
    X(-\Delta) & = & \left(\frac{2}{\omega_{0}}\right) \frac{(\Omega - \Delta + \dot{\imath})v^{\textrm{in}}(-\Delta) + F v^{*\textrm{in}}(\Delta)}{(\Omega + \Delta - \dot{\imath})(\Omega -\Delta + \dot{\imath}) - F^2}\nonumber\\
    & = & \left(\frac{2}{\omega_{0}}\right)\left[ A(-\Delta) \; v^{\textrm{in}}(-\Delta) + B(-\Delta)\; v^{*\textrm{in}}(\Delta)\right] \label{trans}
\end{eqnarray}
A plot of the $|A(\Delta)|^2$ (\emph{cis}-gain) and $|B(\Delta)|^2$ (\emph{trans}-gain), as a function of detuning from the ghost frequency, is shown in Figs. \ref{draw1}(a) and \ref{draw1}(b) for different pump powers $F$. This shows that the maximum gain is realized at zero detuning (at ghost frequency) and not at the natural frequency of the system $\omega_{0}$ in presence of the drive. Internal gain profile at the signal frequency exhibits a \emph{frequency-pulling effect} which comes into play as the drive is increased from zero. This leads to a progressive shift of the spectrum towards $\omega_{g}$ with increase in drive power and leads to maximum parametric gain at the bifurcation threshold. Note that the \emph{trans}-gain does not `see' the natural frequency $\omega_{0}$ and maximal amplification occurs only at the ghost frequency for all drive powers.
\par
\begin{figure}[t]
\includegraphics[width=0.8\textwidth]{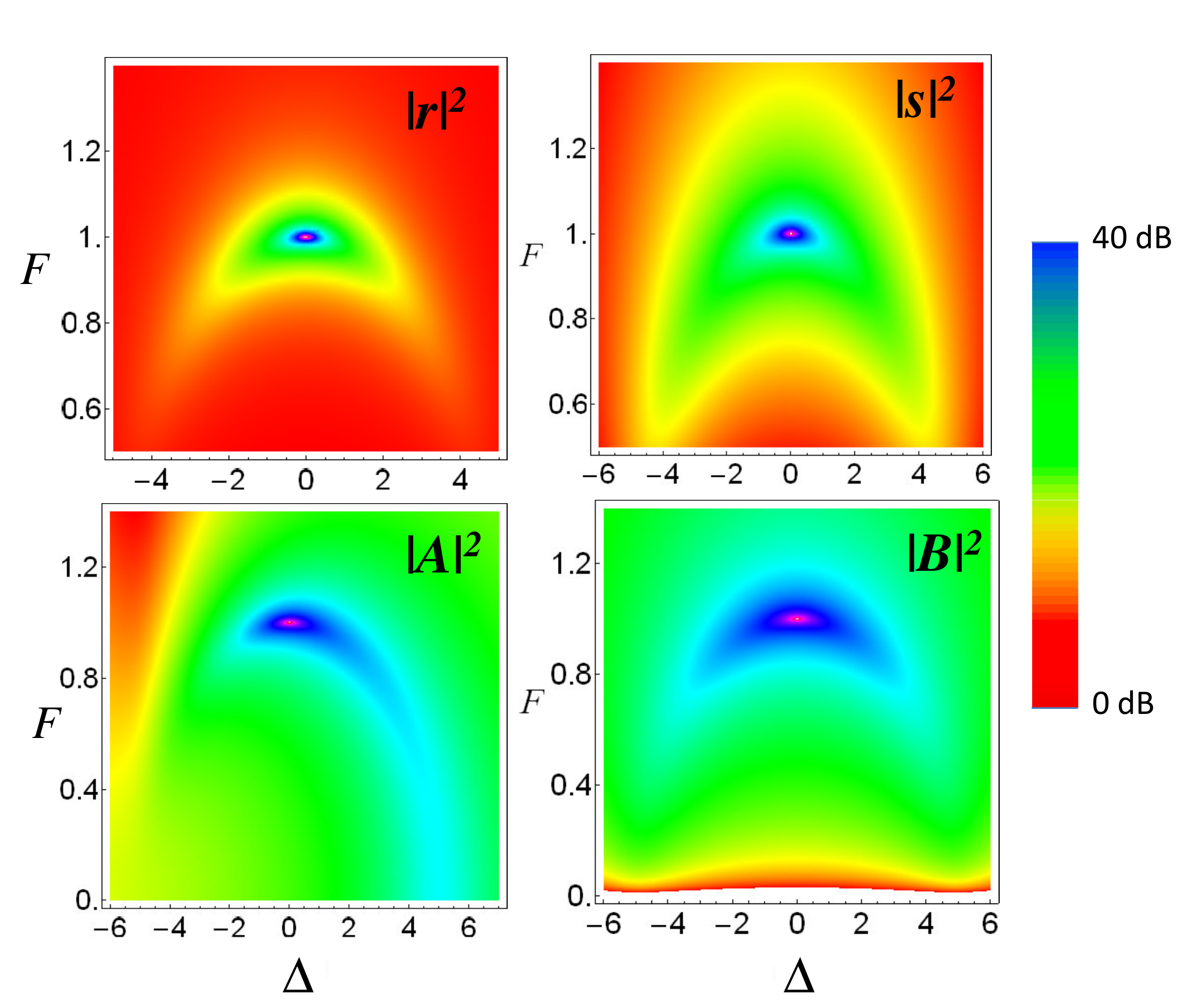}\\
\caption{Upper panels show gain profiles for signal ($|r|^2$) and inter-conversion ($|s|^2$) gains while lower panels show \emph{cis} ($|A|^2$)- and \emph{trans} ($|B|^2$)- gains as a function of reduced drive strength $F$ and signal detuning $\Delta$. Here the value $\Omega = 5$, which is the upper limit of the frequency range in Fig. \ref{draw1}, was used to display the salient features clearly. The plots for $|r|^2$ and $|s|^2$ are symmetric about the ghost frequency corresponding to $\Delta = 0$. The internal trans-gain of the amplifier also exhibits this symmetry. However, the cis-gain is asymmetric since it knows about the relative positions of $\omega_{0}$ and $\omega_{g}$.}\label{density}
\end{figure}

The coefficients $A(\Delta)$ and $B(\Delta)$ crucially decide the gain and noise performance of the amplifier. This follows from the relation between the input and the output fields of the amplifier \begin{center} $v^{\textrm{out}}[\omega] = -\dot{\imath} \omega \varphi[\omega] - v^{\textrm{in}}[\omega]$\end{center} which leads to:
\begin{eqnarray}
\left( \begin{array}{l}
    \hat{a}^{\textrm{out}}[\omega_{S}]\\ \hat{a}^{\textrm{out}}[-\omega_{S}]\\ \hat{a}^{\textrm{out}}[\omega_{I}]\\
    \hat{a}^{\textrm{out}}[-\omega_{I}] \end{array} \right)
    = \left(\begin{array}{cccc}
    e^{\dot{\imath} \alpha}r & 0 & 0 & e^{\dot{\imath} \alpha}s \\
    0 & e^{-\dot{\imath} \alpha}r^{*} & e^{-\dot{\imath} \alpha}s^{*} & 0 \\
    0 & e^{-\dot{\imath} \alpha}s & e^{-\dot{\imath} \alpha}r & 0 \\
    e^{\dot{\imath} \alpha}s^{*} & 0 & 0 & e^{\dot{\imath} \alpha}r^{*}
\end{array} \right)
\left( \begin{array}{l}
        \hat{a}^{\textrm{in}}[\omega_{S}]\\ \hat{a}^{\textrm{in}}[-\omega_{S}]\\ \hat{a}^{\textrm{in}}[\omega_{I}]\\
        \hat{a}^{\textrm{in}}[-\omega_{I}] \end{array} \right)
\end{eqnarray}
where $r = -2\dot{\imath} A - 1$ and $s=-2\dot{\imath} B$. Thus, from Eqns. (\ref{cis}) and (\ref{trans}), we obtain
\begin{eqnarray}
    \boxed{\begin{array}{ccc}
    \displaystyle r = \frac{(1-\Omega^2 + \Delta^2 + F^2) - 2 \dot{\imath} \Omega}
    {|(\Omega - \Delta - \dot{\imath})(\Omega + \Delta + \dot{\imath}) - F^2|} \; \; \; ;
    s = \frac{-2 \dot{\imath} F}{|(\Omega - \Delta - \dot{\imath})(\Omega + \Delta + \dot{\imath}) - F^2|}\\
    \displaystyle \alpha = \arg{\left[\frac{1}{(\Omega - \Delta - \dot{\imath})(\Omega + \Delta + \dot{\imath}) - F^2}\right]}
    \end{array}}
    \label{main8}
\end{eqnarray}
Here we have introduced $\hat{a}^{\textrm{in/out}}$ are the respective field operators for the input and the output ports of the line (cf. Appendix \ref{IOT_app}). It can be verified that the resulting scattering matrix is \emph{symplectic} in nature, i.e. ${^T S} J S = J$ \cite{bergeal-2008} where
\[ J = \left(\begin{array}{cccc}
        0 & 1 & 0 & 0\\
        -1 & 0 & 0 & 0\\
        0 & 0 & 0 & 1\\
        0 & 0 & -1 & 0
        \end{array} \right)
\]
and has a unit determinant. The property of symplecticity ensures the absence of any extraneous degrees of freedom, and hence no missing information (\emph{information preservation}) - a condition necessary for quantum-limited detection \cite{Clerk1, ClerkRMP2}. Moreover, $|r|^2 -|s|^2=1$. Thus we can write
\begin{eqnarray}
   \hat{ a}^{\textrm{out}}[\omega_{S}] = \sqrt{G} \; \hat{a}^{\textrm{in}}[\omega_{S}] + \sqrt{G -1}\;\hat{a}^{\textrm{in}}[-\omega_{I}]
    \label{main9}
\end{eqnarray}
when $\omega_{S}$ and $\omega_{I}$ are very close and phase factors can be ignored. Here $G = |r|^{2}$ is the power gain of the amplifier. A relation of the form shown in Eq. (\ref{main9}) is typical of an amplifier operating at the quantum limit. The internal mode fluctuations of the amplifier are the vacuum fluctuations at the idler port, which are the source of noise added by the amplifier. The efficiency of their conversion to signal frequency at the output port is indicated by the inter-conversion gain (or $|s|^2$). Fig. \ref{density} shows the profile of the signal and inter-conversion gains on a two dimensional color plot, as a function of drive power and detuning. For the sake of comparison, plots for cis- and trans-gains are also shown. The maximum signal and inter-conversion gain profiles are symmetric and indicate maximum gain for both positive (signal) and negative (idler) values of detuning at the ghost frequency and the bifurcation threshold.
\par
Further, we can calculate the output noise spectrum from Eq. (\ref{main9}) as
\begin{eqnarray*}
    \hbar \omega_{S} S_{aa}^{\textrm{out}}[\omega_{S}] = \hbar \omega_{S}  G S_{aa}^{\textrm{in}}[\omega_{S}] + \hbar \omega_{S}  (G - 1) S_{aa}^{\textrm{in}}[\omega_{I}].
\end{eqnarray*}
where $S_{aa}^{\textrm{in/out}}$ represent the input/output photon number spectral densities. This gives the added noise as $ E_{N}^{\textrm{out}} = \hbar \omega_{S}  (G - 1) S_{aa}^{\textrm{in}}[\omega_{I}]$ with $S_{aa}^{\textrm{in}} = \frac{1}{2} \coth \left( \frac{\hbar \omega_{S} }{2 k_{B} T}\right) $ (cf. Appendix \ref{IOT_app}). When referring back to the input, we obtain the noise temperature of the amplifier, $T_{N} = \frac{E_{N}^{\textrm{out}}}{k_{B} G}$, as
\begin{eqnarray}
    \boxed{T_{N} = \frac{\hbar \omega_{S} }{2 k_{B}}}
\end{eqnarray}
for $G \gg 1$ and $T \rightarrow 0$. Thus, like a single pump paramp, the DPA adds half a photon at the signal frequency and approaches arbitrarily close to quantum limited behavior when operated at $k_{B}T \ll \hbar \omega$.
%
%
%ccccccccccccccccccccccccccc
\subsection{Steady State Calculation}
\label{sec_st_stiff}
%ccccccccccccccccccccccccccc
%
%
Self-oscillation is defined as the response of the system in the absence of any input. To compute this for the DPA, we use Eq. (\ref{main7}), under the stiff pump approximation (ignoring the noise back-action factor $\Pi$). We need to reinstate the $\Xi^{2}$ term in $F$ for this calculation as the coupling between $\Xi$ and $F$ is responsible for development of a self-oscillation amplitude in the system. Thus a strictly stiff pump, like that used for derivation of noise spectra in the preceding section, cannot lead to spontaneous emission of finite amplitude. The inclusion of back-action to some zeroth order, i.e. due to self-oscillation itself and \emph{not} due to noise, is crucial for the prediction of this effect.
\par
\begin{figure}
\centering
\includegraphics[width=0.8\textwidth]{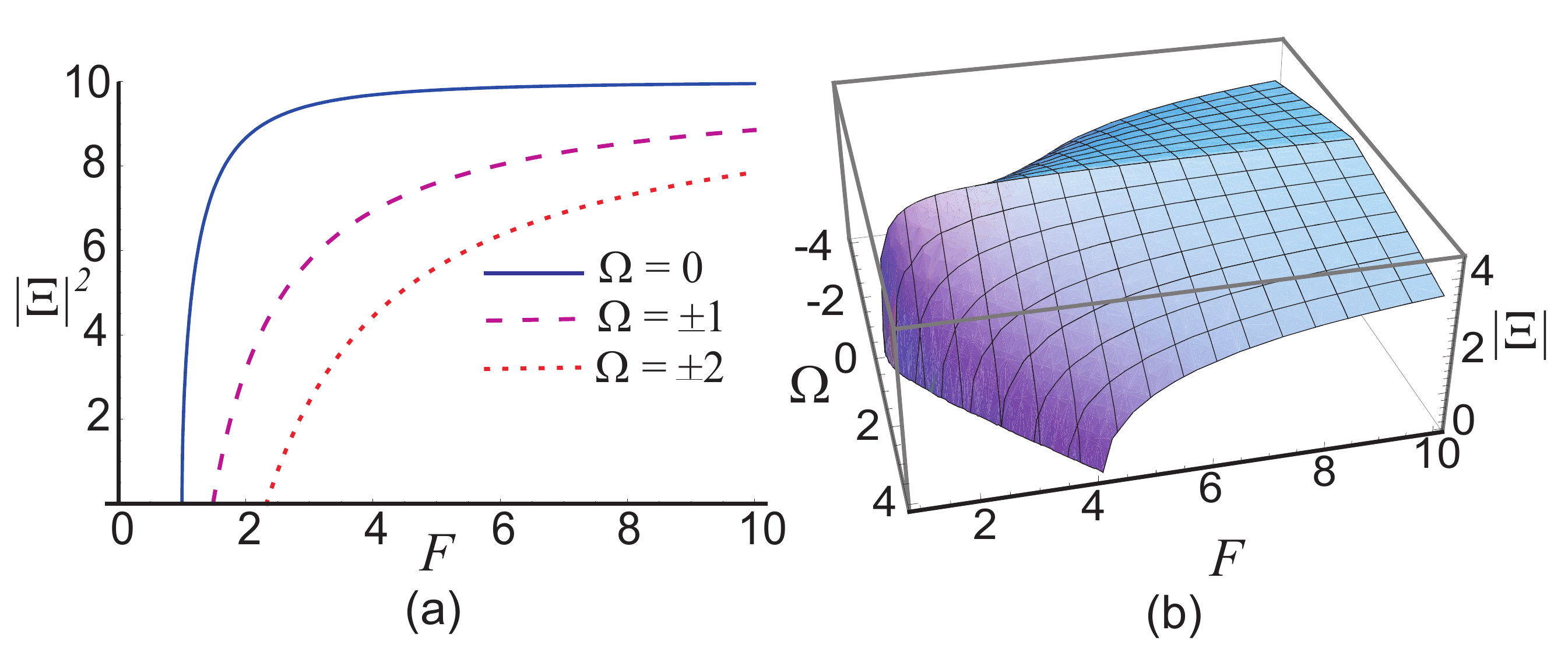}
\caption{Self-oscillation amplitude as a function of drive $F$ and reduced pump detuning $\Omega$. Panel (a) represents cuts of the surface shown in panel (b). Note that self-oscillation develops only beyond a bifurcation threshold $F_{B}$ which has a parabola-like dependence on the reduced pump detuning. The lowest bifurcation value $F_{B} = 1$ is found at zero pump detuning (i.e. $\omega_{g} =\omega_{0}$) and defines the minimum/critical bifurcation power $F_{c}$.}
\label{draw2}
\end{figure}

From Eq. (\ref{main6}), we have
\begin{eqnarray}
    (\Omega -\dot{\imath}) \Xi + \Xi^{*}F \left(-1 + \frac{\gamma}{2}\Xi^2 \right)= 0.
    \label{main10}
\end{eqnarray}
Writing the equation for complex conjugate $\Xi^{*}$ and eliminating $\Xi^{*}$ from both the equations in favor of $\Xi$, we get a cubic equation in $|\Xi|^2$.
\begin{eqnarray}
    (\Omega^2 + 1 - F^2)|\Xi|^2  +  \Omega \gamma F |\Xi|^4 + \frac{1}{4}\gamma^2 F^2 |\Xi|^6 = 0
    \label{stateeqDPA}
\end{eqnarray}
On factoring out the zero solution $|\Xi|^2$ and ignoring the imaginary solution, we get the solution for self-oscillation amplitude as:
\begin{eqnarray}
    \boxed{|\Xi|^2 = 2 \left(\frac{- |\Omega| + \sqrt{F^2-1}}{\gamma F}\right)}
    \label{main11}
\end{eqnarray}
A plot of Eq. (\ref{main11}) (cf. Figs. \ref{draw2}a and \ref{draw2}b) shows that beyond the threshold power $F_{B}$, the system develops a finite amplitude of self-oscillation. This phenomena is reminiscent of \emph{lasing} although here the frequency of self-oscillation is imposed by a combination of the frequencies of the two pumps. This also justifies our simplifying assumption of ignoring the back-action due to $\Xi$ while calculating signal and idler gains, as we see that before bifurcation threshold this approximation is exact.
\par
\begin{figure}
\includegraphics[width=0.8\textwidth]{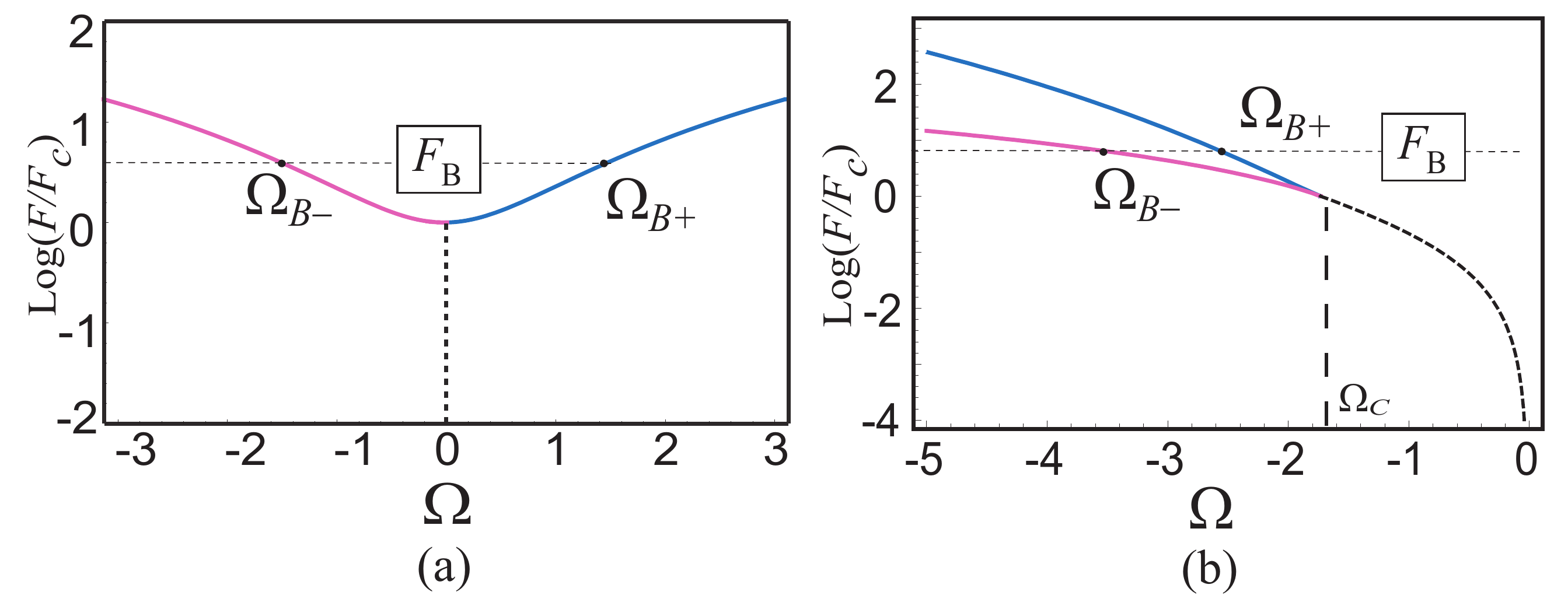}
\caption{(a) Phase diagram of the DPA in absence of noise at the input port. Self-oscillation occurs above the solid line. The vertical dashed line represents the condition for optimal parametric pumping. In this diagram, the bifurcation value $F_{B}$ is a single-valued function of $\Omega$. (b) Corresponding phase diagram of the Josephson Bifurcation Amplifier (JBA), showing the bifurcation curves for upper (blue) and lower (pink) bifurcation points. Here, there exists a critical detuning $\Omega_{c} = \sqrt{3}$ below which the system does not exhibit bifurcation. Below $\Omega_{c}$, the bifurcation power is a double-valued function of $\Omega$. The curved dashed line corresponds to the location of maximum oscillation amplitude. See Fig. \ref{final} (Appendix \ref{JBA_app}) for the behavior of oscillation signal as a function of drive strength in the JBA.}
\label{draw3}
\end{figure}

The phase diagram of the DPA (cf. Fig. \ref{draw3}a) depicting the locus of bifurcation follows from Eq. (\ref{main11}). Since $F_{B} = \sqrt{\Omega^2 +1}$, we see that the phase diagram is symmetric about $\Omega = 0$. Also, the minimum pump power required to make the system self-oscillating ($F_{c}$) is unity, which is realized when system is pumped at resonance ($\omega_{1}+\omega_{2} = 2 \omega_{0}$). This scenario may be contrasted with the phase diagram of the Josephson bifurcation amplifier (JBA) which employs a single pump source (cf. Fig. \ref{draw3}b). The phase diagram of the JBA shows an inherent skewness, which is absent in the DPA. Moreover, the optimal bias point for self-oscillations \emph{does not} coincide with the optimal point of operation ensuring maximum parametric amplification. For the DPA, by contrast, the two points coincide at the global minimum of the $F-\Omega$  plane. Appendix \ref{JBA_app} contains a detailed discussion of the JBA phase diagram. Calculation of explicit transition probabilities between the two stable states resulting in amplification near bifurcation for single pump systems can be found in Ref. 31.
\par
\begin{figure}[h!]
\includegraphics[width=0.5\textwidth]{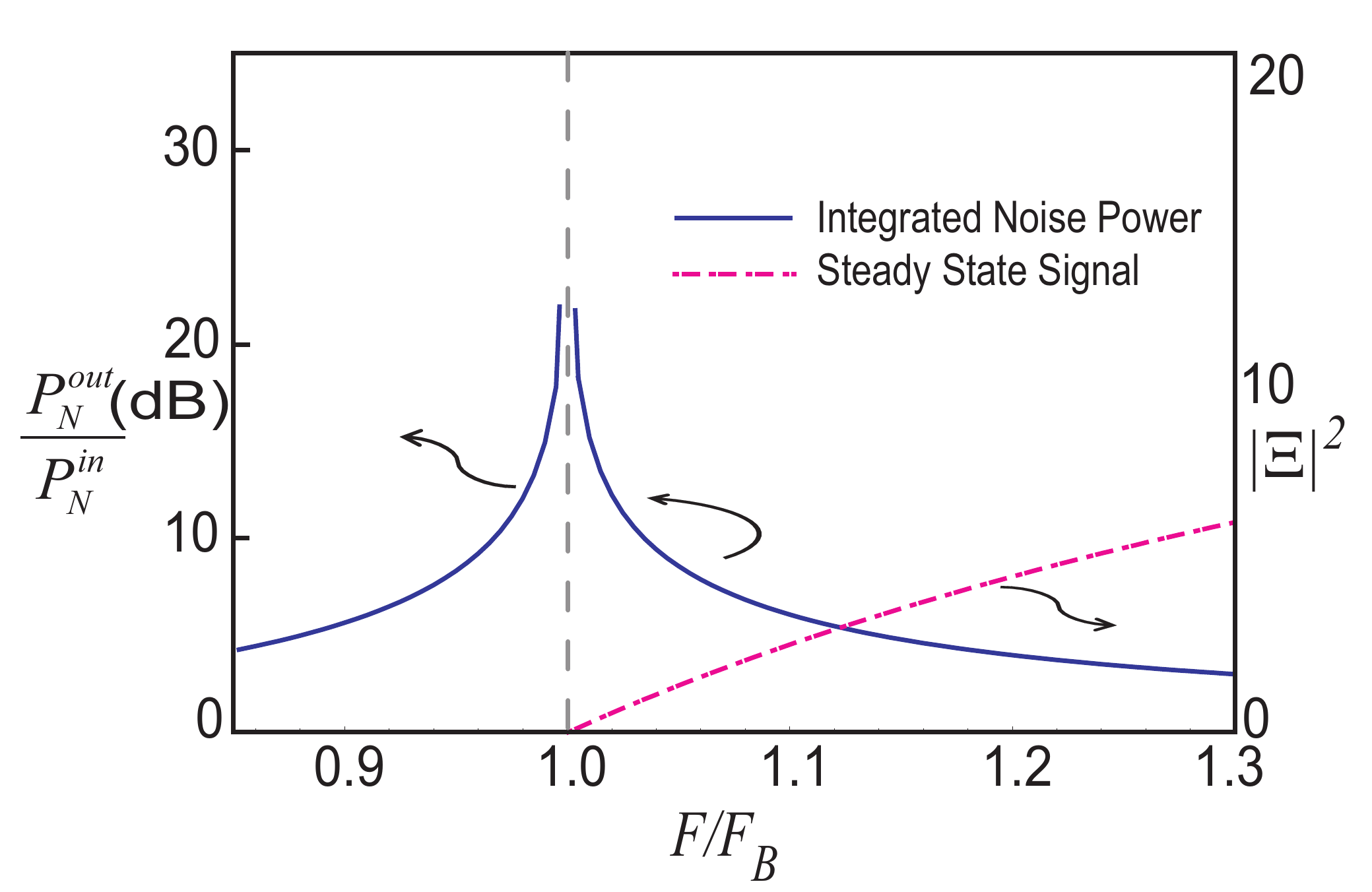}\\
\caption{Integrated noise power gain and self-oscillation amplitude, for the case of stiff pump with $\Omega =1.0$, as a function of reduced pump power. The pump frequencies were $\omega_{1} = 3/4 \omega_{0}, \omega_{2} = 5/4 \omega_{0}$, giving $\gamma \simeq 0.1$.}
\label{draw4}
\end{figure}

It is useful to plot the linear response of the amplifier to wide band noise and classical self-oscillation together, as in Fig. \ref{draw4}. The integrated noise power gain is calculated as
\begin{eqnarray}
    \frac{P_{N}^{\textrm{out}}}{P_{N}^{\textrm{in}}} = \frac{1}{2 \Gamma} \int_{-\Gamma}^{\Gamma}|r[\omega]|^2 d(\omega).
\end{eqnarray}
Fig. \ref{draw4} shows an important result: the threshold power for lasing coincides with the power corresponding to maximum of the integrated noise spectrum. This strongly suggests that the origin of self-oscillation of the system is a result of coherent multi-wave mixing of correlated noise components, which grow around the effective pump frequency $\omega_{g}$ with an increase in the drive. These noise correlations, resulting from parametric interactions, grow stronger as we drive the system near bifurcation threshold and ultimately lead the system to self-oscillate. Thus, beyond bifurcation threshold, there is an additional channel coinciding with the effective pump frequency $\omega_{g}$ in which the system leaks energy.
%
%
%ccccccccccccccccccccccccccccccc
\subsection{Squeezing}
\label{sec_squ_stiff}
%ccccccccccccccccccccccccccccccc
%
%
One of the most attractive applications of parametric amplifiers is the generation of squeezed states of electromagnetic signals \cite{squeezed1, squeezed2}. A squeezed field is one with phase-sensitive quantum fluctuations, which are amplified in one quadrature and deamplified in the other below the quantum vacuum floor. They exhibit strong nonclassical effects which find applications in optical communication systems, interferometry-based gravitational wave detection \cite{0264-9381-23-8-S31} and quantum cryptography \cite{PhysRevA.61.042302}. Squeezing using Josephson parametric devices has already been realized in four-wave mixing configurations \cite{PhysRevLett.56.788, Yurke:87, PhysRevLett.60.764, PhysRevA.39.2519} using single pump. Here we will show the identification of squeezing produced by the two-pump scheme.
\par
\begin{figure}[h!]
\includegraphics[width=0.7\textwidth]{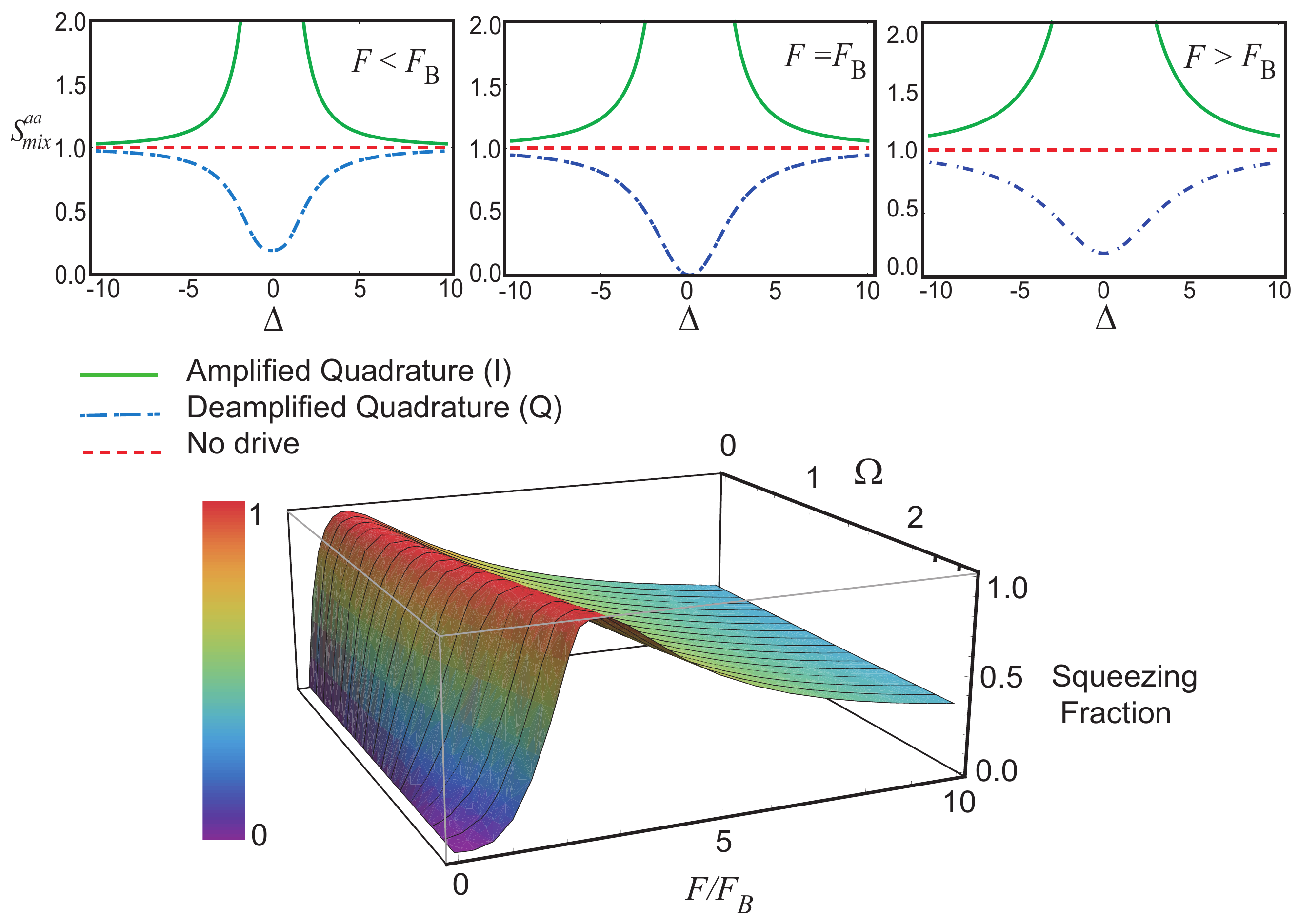}\\
\caption{Squeezing action of the DPA. (\emph{Upper panel}) Deamplified and amplified quadrature power gain for different drive strengths, as a function of reduced signal detuning, with $\Omega = 1$. Maximum squeezing corresponds to the bifurcation threshold ($F = F_{B}$ in Fig. \ref{draw4}), obtained for both peak noise power and onset of self-oscillation. (\emph{Lower panel}) Squeezing fraction = $1 - \sqrt{\frac{I}{Q}}$ increases from 0 (no squeezing) at zero drive to its maximum value 1 (perfect squeezing) at $\Omega$-dependent bifurcation threshold. Beyond the threshold, squeezing decreases.}
\label{squeezed}
\end{figure}

The basic idea is to exploit the correlation between the generated signal and idler photons. The output field of the oscillator is first mixed with a signal at ghost frequency \cite{Vijay}. Then, the homodyne spectrum at the output of the mixer can exhibit sub-quantum noise level for a suitably chosen phase difference ($\theta$) between the original drive and beat signal.
Using the standard quantum optics procedure, the mixing operation in frequency domain can be written as:
\begin{eqnarray}
    \hat{a}^{\textrm{mix}} (\Delta) = \exp(-\dot{\imath}\theta) \hat{a}^{\textrm{out}}(\Delta)
    + \exp(\dot{\imath}\theta) \hat{a}^{\textrm{out}}(-\Delta).
\end{eqnarray}
It can be confirmed that the above relation is identical to that obtained for the output field operator of a two-mode squeezed vacuum state, under the identification $G(=|r|^2)=\cosh^2 t$ in Eq. (\ref{main9}) \cite{bookGerry}. Using the expressions for $\hat{a}^{\textrm{out}}$, coupled with the commutation relations (cf. Appendix \ref{IOT_app}),
\begin{equation*}
    \langle \{\hat{a}^{\textrm{in}}(\Delta), \hat{a}^{\dag \textrm{in}}(\Delta')\}\rangle = S_{aa} \delta (\Delta - \Delta')
\end{equation*}
where $S_{aa}$ is the spectral density of quantum noise level, we get
\begin{eqnarray}
    S^{\textrm{mix}}_{aa} =  & & S_{aa}(|A^{\textrm{out}}(\Delta)|^{2} + |B^{\textrm{out}}(\Delta)|^{2}
    + |A^{\textrm{out}}(-\Delta)|^{2} + |B^{\textrm{out}}(-\Delta)|^{2} \nonumber\\
    & & \;\;\;\;\; + \;\; 2 Re \{ \exp(2\dot{\imath}\theta )(A^{\textrm{out}}(\Delta)B^{\textrm{out}}(-\Delta)
    + A^{\textrm{out}}(-\Delta)B^{\textrm{out}}(\Delta)) \} )
\end{eqnarray}
Here $A^{out}= -2 \dot{\imath} A - 1$ and $B^{out} = -2 \dot{\imath} B$ with $A$ and $B$ defined by Eqns. (\ref{cis}) and (\ref{trans}). Thus, $S^{\textrm{mix}}_{aa}$ is reduced below $S_{aa}$ for a particular value of $\theta$ and shows a maximum for the orthogonal quadrature. The squeezing spectra for the amplified and deamplified quadratures are shown in Fig. \ref{squeezed}.
\par
It can be seen that maximum squeezing is attained at zero detuning ($\omega_{S} = \omega_{g}$) and the bifurcation threshold corresponding to a given $\Omega$. It may be noted that perfect squeezing at bifurcation is obtained as a result of our stiff pump analysis. The more sophisticated soft pump analysis would predict a squeezing fraction less than 100\% even at the bifurcation.
%
%
%cccccccccccccccccccccccccccccccc
\section{Soft Pump: Inclusion of Back-action Corrections}
\label{sec_ba}
%cccccccccccccccccccccccccccccccc
%
%
To incorporate the effect of back-action on the dynamics, we use the full expression given in Eq. (\ref{main7}) for the pump amplitudes to calculate the response functions of interest. We adopt a mean-field approach in this calculation. Various levels of coupling in the system, manifest in Eqns. (\ref{main1}) to (\ref{main6}), necessitate such a self-consistent analysis of the problem at hand. To make a reasonable start, we pick up one of the strings of this self-consistent loop - signal and idler amplitudes, and calculate them treating $\Pi$ as a parameter of the problem to be evaluated later.
%
%ccccccccccccccccccccccccccccccc
\subsection{Corrections to signal and idler amplitudes}
\label{si_corr}
%ccccccccccccccccccccccccccccccc
%
%
Using the same method as in the previous section - this time using the full expression of pump amplitudes from Eq.(\ref{main7}) - we solve for signal and idler amplitudes and obtain
\begin{subequations}
\begin{align}
    X[\omega_{S}] & = \left(\frac{2}{\omega_{0}}\right) \frac{(\Omega + \Delta +\dot{\imath}) v^{\textrm{in}}[\omega_{S}]
    +  F \left(1 - \gamma(\Pi + \frac{1}{2} \Xi^2) \right) v^{\textrm{in}}[-\omega_{I}]}
    {(\Omega - \Delta - \dot{\imath})(\Omega + \Delta+ \dot{\imath})
    - F^2 \left(1 - \gamma(\Pi + \frac{1}{2} \Xi^2) \right)\left(1 - \gamma(\Pi^{*} + \frac{1}{2} \Xi^{*2}) \right)}
    \label{main12fir} \nonumber\\
    & = \left(\frac{2}{\omega_{0}}\right) \left[\tilde{A}[\omega_{S}] \; v^{\textrm{in}}[\omega_{S}] + \tilde{B}[\omega_{S}]\;v^{\textrm{in}}[-\omega_{I}] \right]\\
    \nonumber\\
    X[\omega_{I}] & =  \left(\frac{2}{\omega_{0}}\right) \frac{(\Omega - \Delta + \dot{\imath})v^{\textrm{in}}[\omega_{I}]
    + F \left(1 - \gamma(\Pi + \frac{1}{2} \Xi^{2}) \right) v^{\textrm{in}}[-\omega_{S}]}
    {(\Omega + \Delta- \dot{\imath})(\Omega -\Delta+ \dot{\imath})
    - F^2\left(1 - \gamma(\Pi + \frac{1}{2} \Xi^2) \right)\left(1 - \gamma(\Pi^{*} + \frac{1}{2} \Xi^{*2}) \right)}
    \label{main12sec}\nonumber\\
    & = \left(\frac{2}{\omega_{0}}\right) \left[\tilde{A}[\omega_{I}] \; v^{\textrm{in}}[\omega_{I}] + \tilde{B}[\omega_{I}]\; v^{\textrm{in}}[-\omega_{S}]\right].
\end{align}
\end{subequations}
where $\tilde{A}$ and $\tilde{B}$ are the corresponding coefficients for a soft pump, analogous to those in Eqns. (\ref{cis}) and (\ref{trans}). It is convenient, both in terms of notation and analysis, to define a \emph{back-action corrected drive power}, $F_{BA}$, given by
\begin{eqnarray}
    F_{BA}= F \left[1 - \gamma\left(\Pi + \frac{1}{2} \Xi^2\right)\right].
    \label{force}
\end{eqnarray}
Note that $F_{BA}$, unlike F, is complex and depends self-consistently on $X[\omega]$ through an integral ($\Pi$).
\par
As before, we calculate the signal and inter-conversion gains using $v^{\textrm{out}}[\omega] = -\dot{\imath} \omega X[\omega] - v^{\textrm{in}}[\omega]$ in terms of this new drive strength. The resulting scattering matrix S, defined by $a^{\textrm{out}}[\omega] = S a^{\textrm{in}}[\omega]$ is again symplectic despite the complex nature of $F_{BA}$. This shows that the basic tenet of \emph{information preservation} requested during an amplification process is fulfilled even in the presence of back-action. The scattering matrix of the amplifier is evaluated to be
\begin{equation*}
S = \left(\begin{array}{cccc}
    e^{\dot{\imath}\alpha}r & 0 & 0 &  e^{\dot{\imath}(\alpha + \delta)} s \\
    0 & e^{-\dot{\imath}\alpha} r^{*} & e^{-\dot{\imath}(\alpha + \delta)} s^{*}& 0 \\
    0 & e^{-\dot{\imath}(\alpha + \delta)}s &  e^{-\dot{\imath}\alpha}r & 0 \\
    e^{\dot{\imath}(\alpha + \delta)} s^{*} & 0 & 0 & e^{\dot{\imath}\alpha}r^{*}
\end{array}\right)
\end{equation*}
with
\begin{eqnarray}
\boxed{\begin{array}{ccl}
    r & = & \displaystyle {\frac{[1-\Omega^2 + \Delta^2 + |F_{BA}|^{2}] - 2 \dot{\imath} \Omega}
    {|(\Omega - \Delta - \dot{\imath})(\Omega + \Delta + \dot{\imath})-|F_{BA}|^2|}} \\
    s & = & \displaystyle {\frac{-2 \dot{\imath} F_{BA}}
    {|(\Omega - \Delta - \dot{\imath})(\Omega + \Delta + \dot{\imath})- |F_{BA}|^2|}} \\
    \alpha & = & \displaystyle {\arg\left[\frac{1}{(\Omega - \Delta - \dot{\imath})(\Omega + \Delta + \dot{\imath})-|F_{BA}|^2} \right]}; \;\;
    \delta = \arg[F_{BA}].
      \end{array}}
      \label{grp}
\end{eqnarray}
As before (cf. Section \ref{sec_pert_stiff}), $|r|^2 -|s|^2 = 1$. Note that the expressions in (\ref{grp}) reduce to those of Eq. (\ref{main8}) when $F_{BA} = F$. The presence of the additional phase factor $\delta$, which is a function of pump detuning $\Omega$, can be understood by noting that  $\Pi$ is a complex quantity in general.
%
%
%ccccccccccccccccccccccccccccccc
\subsection{Corrections to Steady State Signal}
%ccccccccccccccccccccccccccccccc
%
%
We recalculate the steady state response (i.e. zero input response) of the system, using the back-action corrected drive strength. Eq. (\ref{main6}) gives
\begin{eqnarray}
    (\Omega -\dot{\imath}) \Xi + \chi \Xi^{*} F \left[-1 + \gamma\left(\Pi + \frac{1}{2}\Xi^2 \right)\right]= 0.
    \label{main13}
\end{eqnarray}
On solving this equation, as in section (\ref{sec_st_stiff}), we get the self-oscillation amplitude as:
\begin{eqnarray}
    \boxed{|\Xi|^2 = 2 \left(\frac{- |\Omega| + \sqrt{|F_{BA}|^2-1}}{\gamma F} \right)}.
    \label{main14}
\end{eqnarray}
It is instructive to see that Eq. (\ref{main14}) reduces to Eq. (\ref{main11}), in absence of back-action i.e. $F_{BA} \longmapsto F$.
%
%
%cccccccccccccccccccccccccccccccc
\section{Calculation of the Noise Back-action Factor $\Pi$}
\label{corr_calc}
%cccccccccccccccccccccccccccccccc
%
%
We will use the results derived in section \ref{si_corr} to calculate the noise back-action factor $\Pi$. Using Eqns. (\ref{main12fir}) and (\ref{main12sec}) in Eq. (\ref{main5}), under mean field
\begin{eqnarray*}
    \Pi & = & \left\langle \int_{-\infty}^{\infty} d\omega_{a} \int_{-\infty}^{\infty} d\omega_{b} X[\omega_{a}] X[\omega_{b}]
    \delta(\omega_{a}+\omega_{b}- 2 \omega_{g})\right\rangle \\
    & = & \left(\frac{2}{\omega_{0}}\right)^2 \left\langle \int_{-\infty}^{\infty} d\omega_{a}\int_{\omega_{1}}^{\omega_{2}} d\omega_{b}
    \{A[\omega_{a}] \; v^{\textrm{in}}[\omega_{a}] + B[\omega_{a}]\; v^{\textrm{in}}[\omega_{a} - 2 \omega_{g}]\} \times \right. \\
    & & \;\;\;\;\;\;\;\;\;\;\;\;\;\;\;\;\;\;\;\;\; \left. \{A[\omega_{b}] \; v^{\textrm{in}}[\omega_{b}] + B[\omega_{b}]\; v^{\textrm{in}}[\omega_{b} - 2 \omega_{g}]\}\delta(\omega_{a}+\omega_{b}- 2 \omega_{g}) \right\rangle
\end{eqnarray*}
Exploiting the fact that $\langle V^{\textrm{in}}[\omega_{a}] \cdot V^{\textrm{in}} [\omega_{b}]\rangle = S_{vv}[\omega_{a}]\delta(\omega_{a} + \omega_{b})$ (cf. Appendix \ref{IOT_app}) and recalling that $v^{\textrm{in}} = \frac{V^{\textrm{in}}}{\Phi_{0}/2 \pi}$, only two terms out of the above four survive giving
\begin{equation}
    \Pi = \left( \frac{4 \pi}{\Phi_{0}\omega_{0}}\right)^2 \int_{-\infty}^{\infty} d\omega_{a}
    \{A[\omega_{a}]B[\omega_{b}] +   B[\omega_{a}]A[\omega_{b}]\} S_{vv}[\omega_{a}] \label{main15}.
\end{equation}
Using
\begin{equation*}
    S_{vv}[\omega_{a}] = \frac{Z_{c} \hbar \omega_{a}}{4}\left[ \coth \left(\frac{\hbar\omega}{2 k_{B} T}\right) + 1\right]
    \xrightarrow{T \gg \frac{\hbar\omega}{2 k_{B}}} \frac{Z_{c} k_{B} T }{2}
\end{equation*}
in Eq. (\ref{main15}), we obtain
\begin{subequations}
\begin{align}
    \Pi &= \Theta_\textrm{eff} \int_{-\infty}^{\infty} d(\Delta) \mathcal{P}(\Omega, \Delta, F) \label{fifth}\\
    \lefteqn{\textrm{with}}\hspace{2in} \nonumber\\
    \mathcal{P}(\Omega, \Delta, F) &= \frac{2 (\Omega + \dot{\imath})  F_{BA} }
     {\{(\Omega - \Delta + \dot{\imath})(\Omega + \Delta - \dot{\imath}) - F_{BA}^2 \}
     \{(\Omega - \Delta - \dot{\imath})(\Omega + \Delta + \dot{\imath}) - F_{BA}^2 \}} ; \;\;\;\; \label{fiftha}\\
     \Theta_\textrm{eff} &= \frac{k_{B}T}{E_{J}}.\;\; \label{sixth}
\end{align}
\end{subequations}
\par
Here we have used $ E_{J} = \frac{\Phi_{0}^2}{4 \pi^2 L}, \; \Delta =\frac{\omega_{a} -\omega_{g}}{\Gamma},\; Z_{c}= R, \; \Gamma = \frac{1}{2 R C}$.
\par
\begin{figure}[h!]
\includegraphics[width=0.5\textwidth]{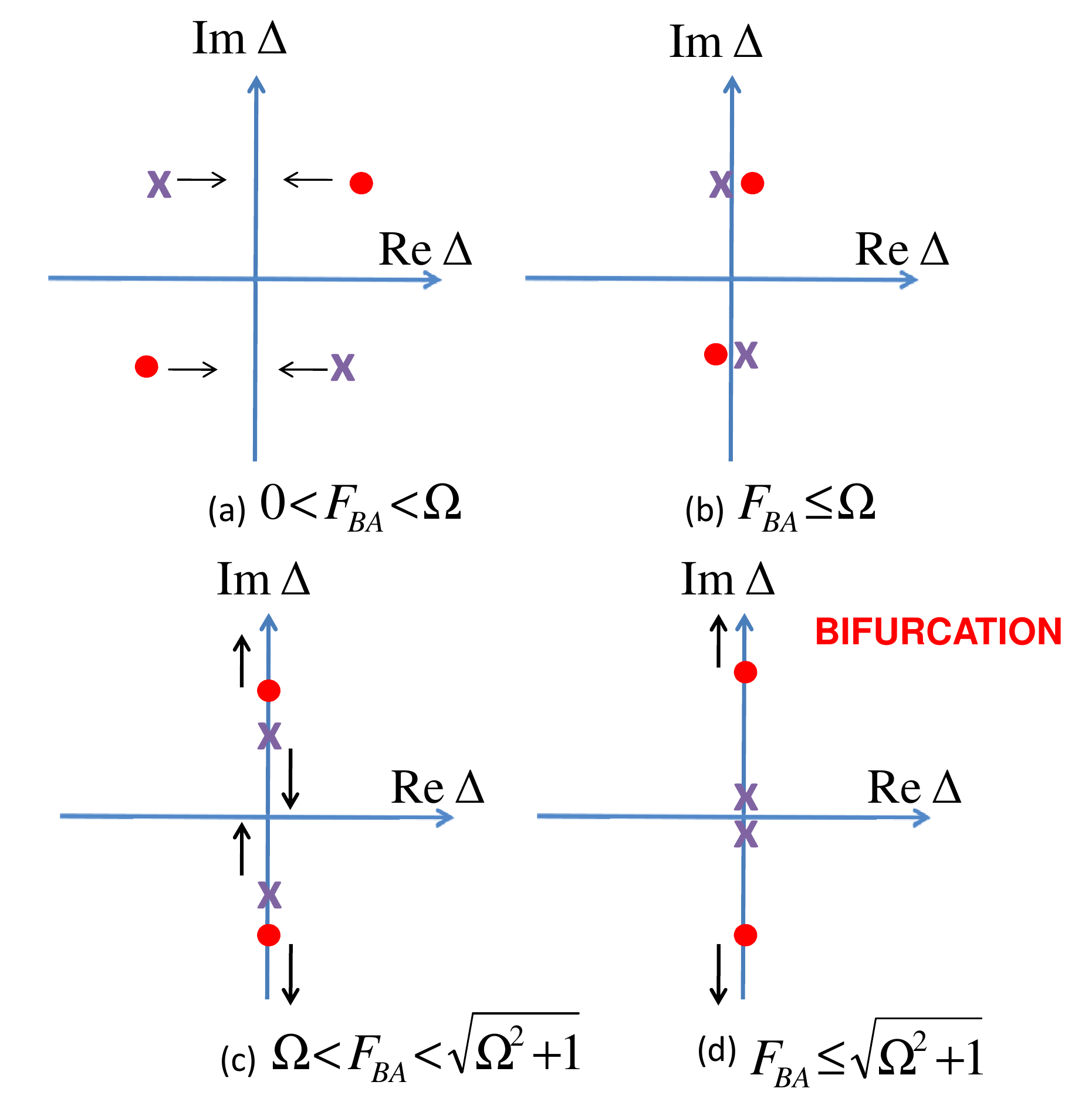}\\
\caption{Poles of the noise back-action factor $\Pi$ in the complex plane of reduced signal frequency $\Delta$. They move as the drive strength is changed. The poles in panel (a) before the bifurcation threshold move towards the imaginary axis. In panel (b), they hit the imaginary axis as $F_{BA}$ takes the value $\Omega$. As $F_{BA}$ is further increased, in panel (c), they move along imaginary axis in two opposite directions. Finally, in panel (d), one pair of poles hits the real axis as $F_{BA}$ attains the bifurcation threshold.}\label{poles}
\end{figure}

To evaluate $\Pi$ in Eq. (\ref{fifth}), we use the method of residues. We find that the denominator of $\mathcal{P}$ has four poles in the complex plane of reduced frequency $\Delta$ at $(-\dot{\imath} \pm \sqrt{\Omega^2 -F_{BA}^2})$ and corresponding complex conjugates - giving one pole in each quadrant (Fig. \ref{poles}). To respect causality we need to take into account, at any given drive power, only the contribution of the poles in the upper half complex plane.  We evaluate the residue at each of the relevant poles and sum them up to get the value of complex-valued integral in Eq. (\ref{fifth}) and obtain
\begin{eqnarray}
\boxed{\Pi = \left \{\begin{array}{ccc}
                 \displaystyle \eta \pi \frac{ (\Omega + \dot{\imath})F_{BA}}
                 {1 -F_{BA}^2 + \Omega^2} \;\;  &   F_{BA} < \sqrt{\Omega^2 + 1}
                 \\
                 \displaystyle \eta \pi \frac{ (\Omega + \dot{\imath})F_{BA}}
                 {(-1 + F_{BA}^2 -\Omega^2 )\sqrt{F_{BA}^2 - \Omega^2}}   \;\;&
                 F_{BA} \geq \sqrt{\Omega^2 + 1}
             \end{array}
\right.}
\label{main16}
\end{eqnarray}
\par
The expressions for $\Pi$ are continuous at the bifurcation threshold but have different slopes and qualitative dependence on $F_{BA}$. This indicates that the response functions calculated using $\Pi$, though continuous may not be symmetric across threshold power as in the case of stiff pump. Also introduced is a new parameter
\begin{equation}
    \eta = \gamma \cdot \Theta_\textrm{eff}
    \label{prefactor}
\end{equation}
which occurs as a pre-factor of $\Pi$ in various gain coefficients and incorporates the effect of both non-linearity($\gamma \propto \lambda$) and effective temperature of the input ($\Theta_\textrm{eff} \propto T$). This serves to restrict the parameter space of the problem and can be identified as a kind of \emph{back-action index}. Henceforth, we will define the system parameters in terms of $\eta$.
\par
\begin{figure}[h!]
\includegraphics[width=0.3\textwidth]{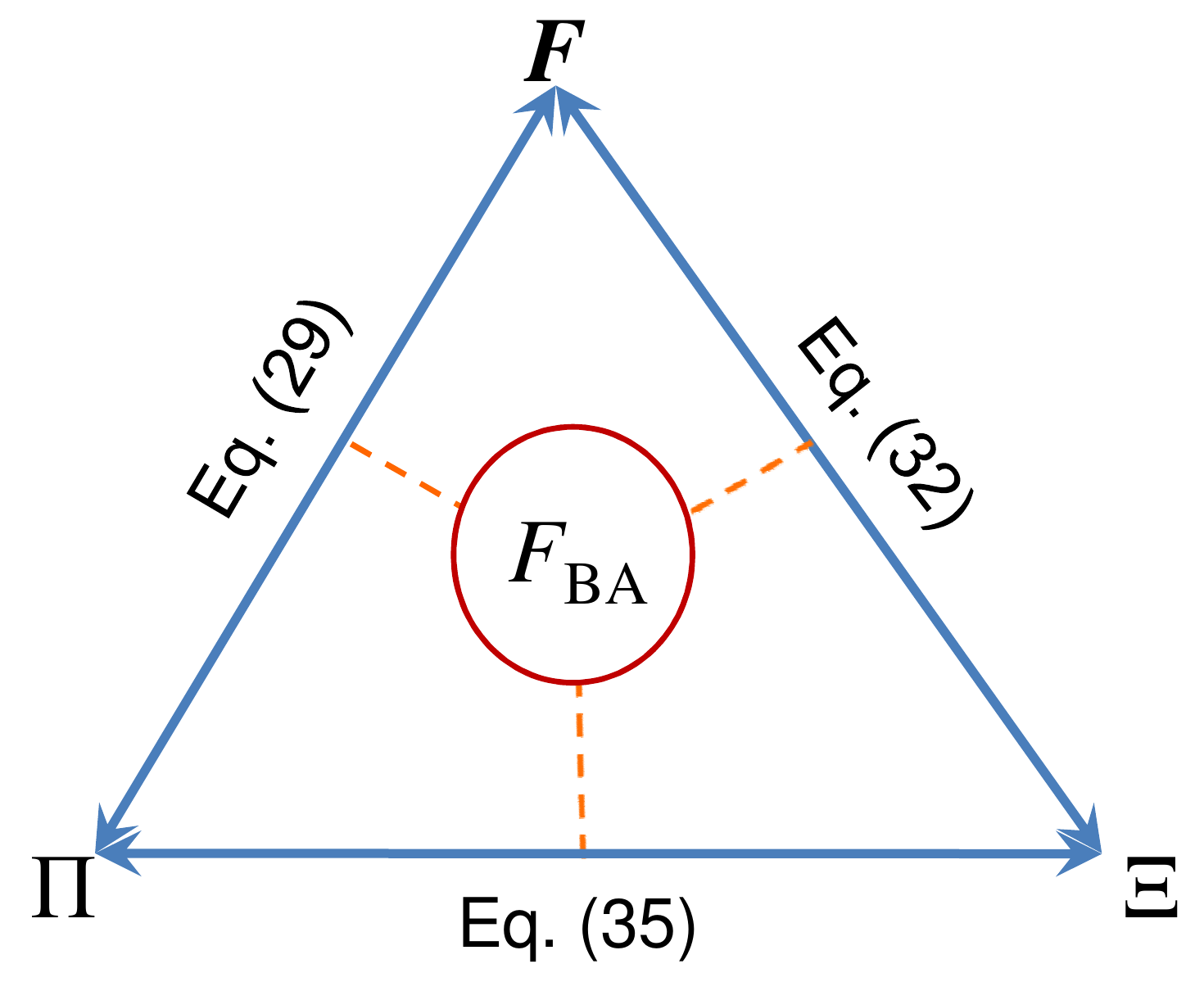}\\
\caption{Schematic diagram displaying the structure of the set of Eqns. (\ref{force}), (\ref{main14}) and (\ref{main16}). The drive strength $F$, self-oscillation amplitude $\Xi$ and noise back-action factor $\Pi$, all involve back-action corrected pump strength $F_{BA}$ and are interdependent .}
\label{self_consistency}
\end{figure}

As evident from Eq. (\ref{main16}), $\Pi$ depends on $F_{BA}$, which itself depends on $\Pi$. Such a relation is a natural consequence of a mean-field approach. Also, it should be noted that as the poles depend on this corrected value of drive power. We need to fully appreciate the three-way coupling reflected by Eqns. (\ref{force}), (\ref{main14}) and (\ref{main16}) before embarking on a solution. Both the complexity and strength of the following analysis lies in the self-consistent treatment encompassing all the three players of the game. In these three equations, we identify $F_{BA}$ as a convenient ``slave'' parameter (Fig. \ref{self_consistency}) and solve the system of equations parametrically.
\par
It may be noted that the self-oscillation term is especially important once the bifurcation threshold is reached and needs to be incorporated to get the correct value of back-action on drive strength. However in the expressions for signal and inter-conversion gains - $|r|^{2}$ and $|s|^{2}$  (Eq. \ref{grp})- only terms like $|F_{BA}|^2$ appear. Due to this, only the real and absolute values of $\Pi$ and $\Xi^2$ are important. Exploiting this fact, we have used the expression obtained for the self-oscillation signal as obtained in Eq. (\ref{main14}), to account for the back-action of self oscillation signal on drive strength ($\Xi^2$ term in Eq. (\ref{force})).
\par
\begin{figure}[h!]
\includegraphics[width=0.5\textwidth]{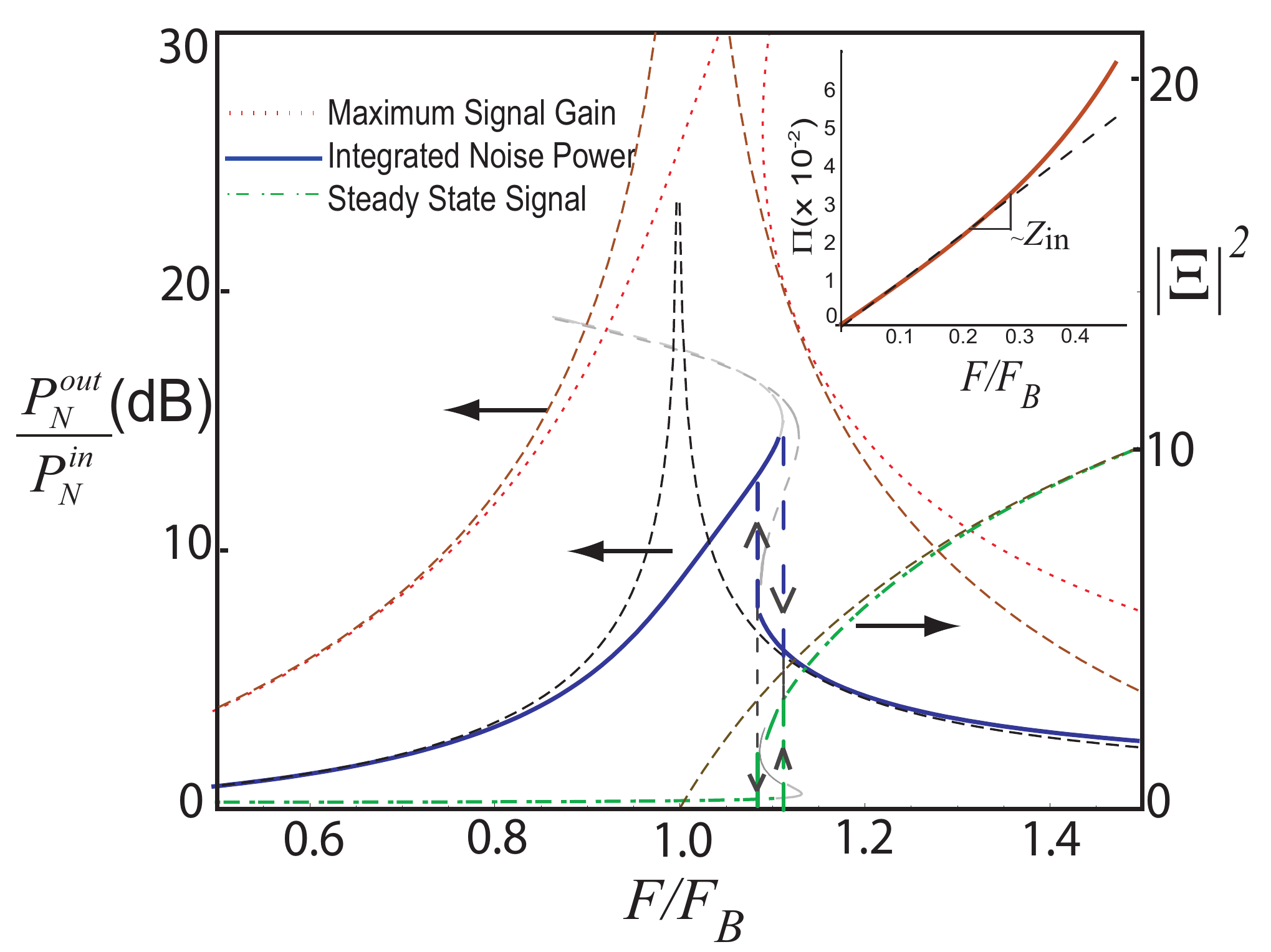}\\
\caption{Maximum signal gain (dotted curve), integrated noise power gain (solid curve), self-oscillation component (dotted-dashed curve) as a function of reduced pump power for $\Omega = 0.5$ and $\eta = 3 \times 10^{-3}$ (values corresponding to $|\lambda| = 1/6$ and $\Theta_{eff} = 3\%$). The dashed curves show the respective stiff pump responses. Back-action manifests itself through a pronounced reentrant behavior and a shift of the bifurcation threshold to higher powers. The inset shows the noise back-action factor $\Pi$ as a function of $F$ evaluated for the same system parameters. The slope represents the impedance seen looking into the pump port. This impedance acquires power dependence as bifurcation threshold is approached.}
\label{draw6}
\end{figure}

In Fig. \ref{draw6}, we show parametric plots of both the integrated noise power gain and self-oscillation amplitude of the system, obtained after incorporating the effect of back-action.
The response functions clearly show that the system exhibits `hysteresis' as indicated by the reentrant gain curves: gain is a multiple valued function of the control parameter (drive strength here). Another consequence of the incorporation of back-action is the reduction of the effective drive power which shifts the threshold power corresponding to the maximum noise gain, as well as the onset of self-oscillation, to higher values. Thus, the behavior of system in the presence of back-action is markedly different from that obtained in the stiff pump approximation (Fig. \ref{draw4}).
\par
\begin{figure}
\centering
\includegraphics[width=\textwidth]{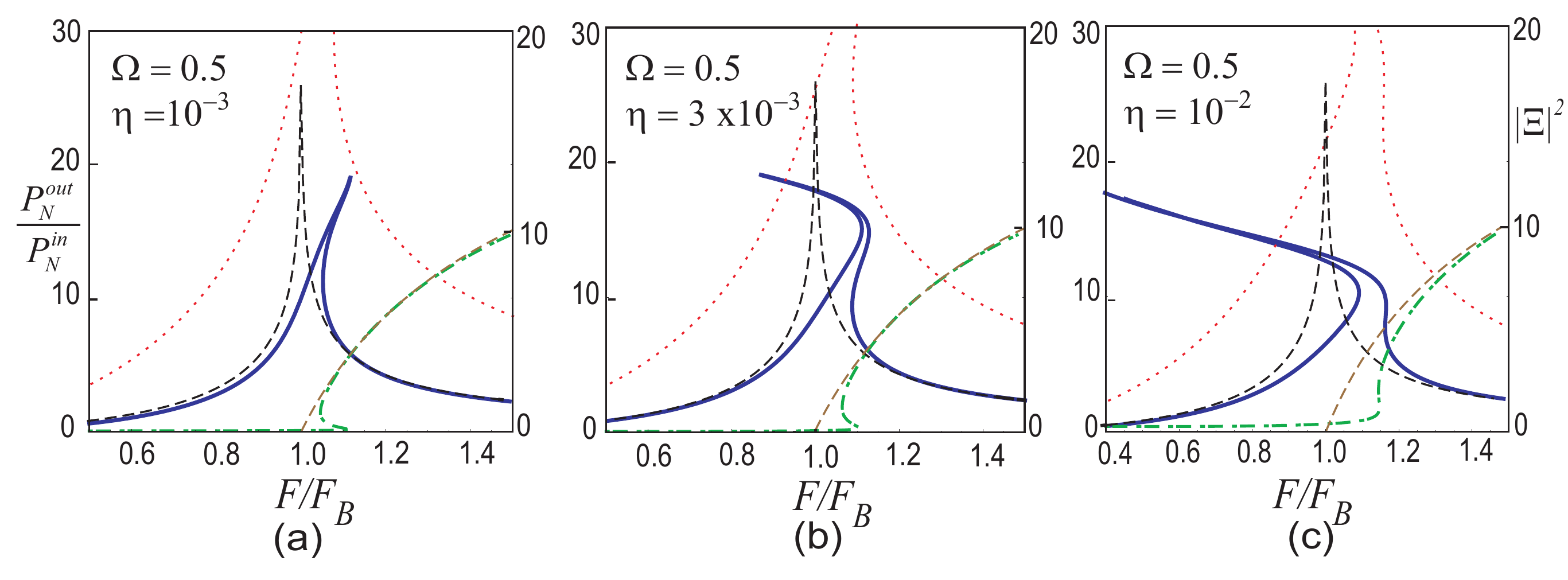}
\caption{Integrated noise power gain in dB (solid curve) and self-oscillation (dotted-dashed curve) responses, at $\Omega = 0.5$, for different values of temperature (a) $\eta = 10^{-3}$ (b) $\eta = 3\times10^{-3}$ and (c) $\eta = 10^{-2}$. The noise back-action factor scales linearly with temperature of the noise at the input and hence its effect becomes more pronounced at higher temperatures. Curves above reflect this through a higher fold-over and increase in threshold power with temperature. The dashed curves representing the behavior of respective stiff pump response functions are included for comparison.}
\label{draw7}
\end{figure}

We now study the effect of various system parameters on the back-action induced corrections. An increase in effective temperature is seen to cause a greater back-action correction. This is depicted in the noise gain and self-oscillation gain curves plotted in Fig. \ref{draw7} for different values of parameter $\eta$, which varies linearly with effective temperature of the input port $k_{B} T /E_{J}$ (cf. Eq. (\ref{prefactor})).
\begin{figure}
\centering
\includegraphics[width=\textwidth]{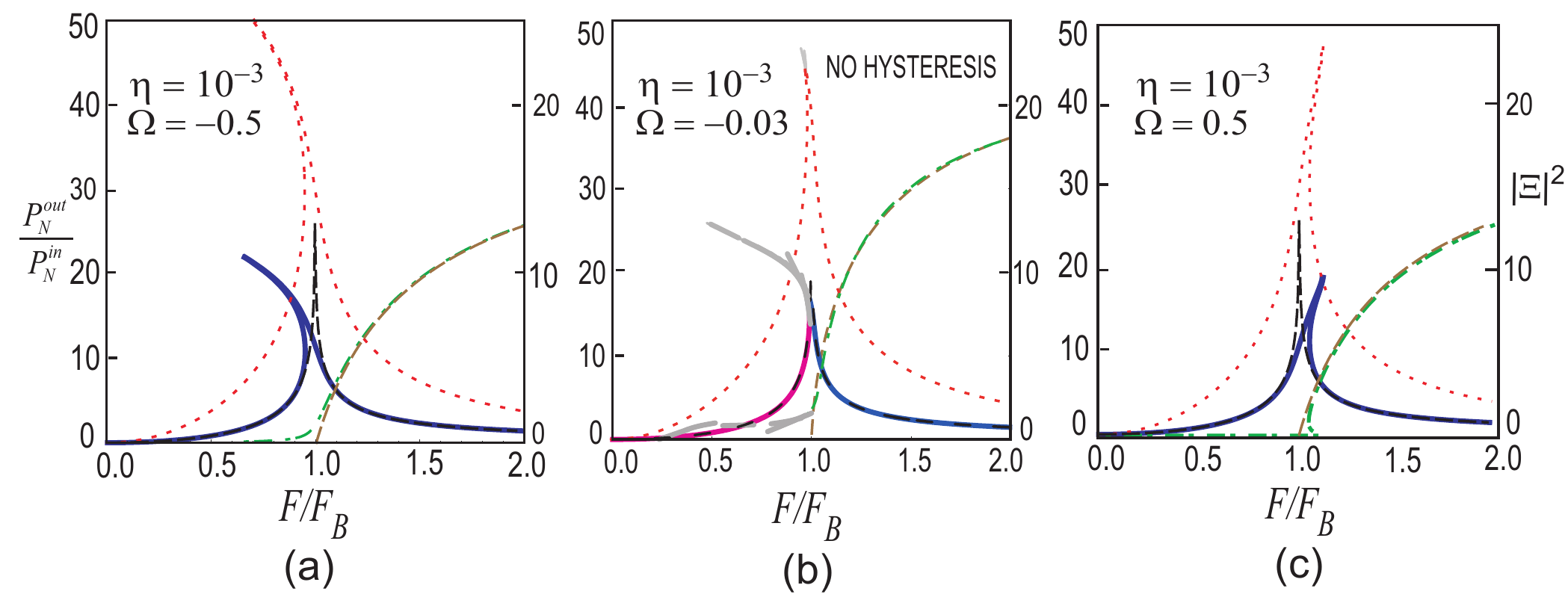}
\caption{This plot shows the dependence of system response on pump detuning $\Omega$ in the presence of back-action  for $\eta = 10^{-3}$. Back-action manifests itself in the form of hysteresis whose range increases as we move away from a renormalized resonant frequency of the oscillator, whose value corresponds here to $\Omega_{0} = -0.03$.  This value, at which hysteresis disappears, is offset from the stiff pump lowest bifurcation value $\Omega = 0.0$. As in Fig. \ref{draw7}, the dashed curves representing the respective stiff pump behavior are included for comparison.}
\label{draw8}
\end{figure}
\par
\begin{figure}
\centering
\includegraphics[width=0.4\textwidth]{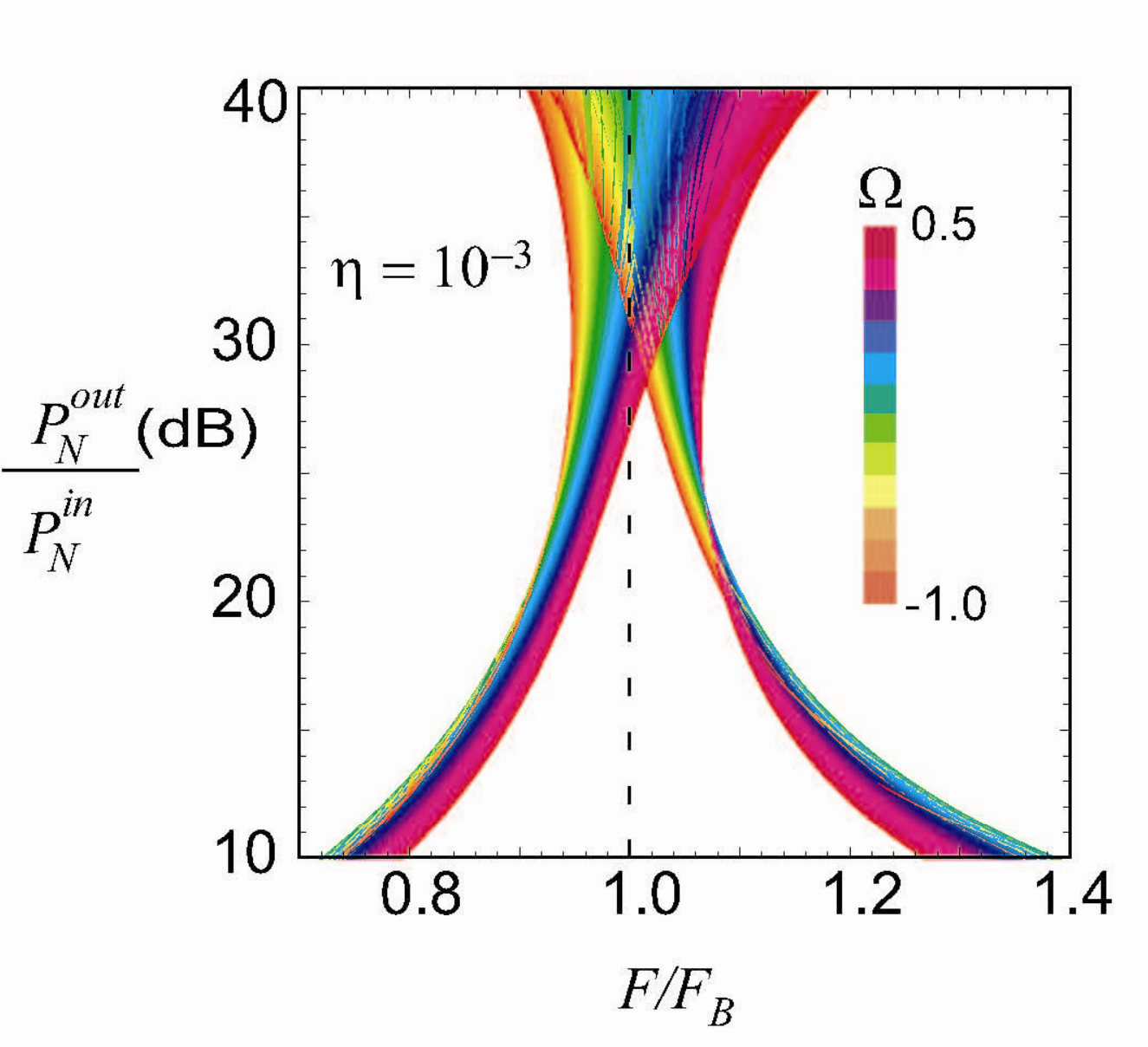}
\caption{Variation of hysteresis due to back-action in integrated noise power gain as a function of pump frequency. The direction of back-bending of the response curves reverses as we move from negative to positive detuning.}
\label{draw9}
\end{figure}

Another parameter of interest in the problem is the reduced pump detuning $\Omega$. It is seen that back-action pushes the bifurcation threshold powers to higher values as $\Omega$ is increased or as we move away from the resonant frequency $\omega_{0}$. Besides an increase in threshold power, the extent of fold-over of gain curves as a function of drive strength also varies as we vary $\Omega$ (Fig. \ref{draw9}). This is depicted in Fig. \ref{draw8}. An important ramification of this effect is that $\Omega = 0 (\omega_{g} = \omega_{0})$ does \emph{not} correspond to resonance any longer, as in the case of stiff pump. This is a consequence of the fact that the resonant frequency of the system $\omega_{0}$ (defined as  $1/\sqrt{L_{J}C_{J}}$ ) itself changes in the presence of a drive, due to back-action. This indicates that for some $|\Omega_{0}| \simeq 0.03$, the system simulates stiff pump behavior and the corresponding $\omega_{0} = \omega_{g}+ \Omega_{0} \Gamma$ corresponds to a \emph{corrected} resonance frequency. Thus, by tuning $\Omega$ close to the new resonance, hysteresis can be eliminated. However, to realize maximum parametric gain at optimal powers \emph{and} relieve the system of hysteresis we need to tune both the parameters F \emph{and} $\Omega$ to ensure the operation of the amplifier near the optimal point in the phase space.
\par
\begin{figure}
\includegraphics[width=0.4\textwidth]{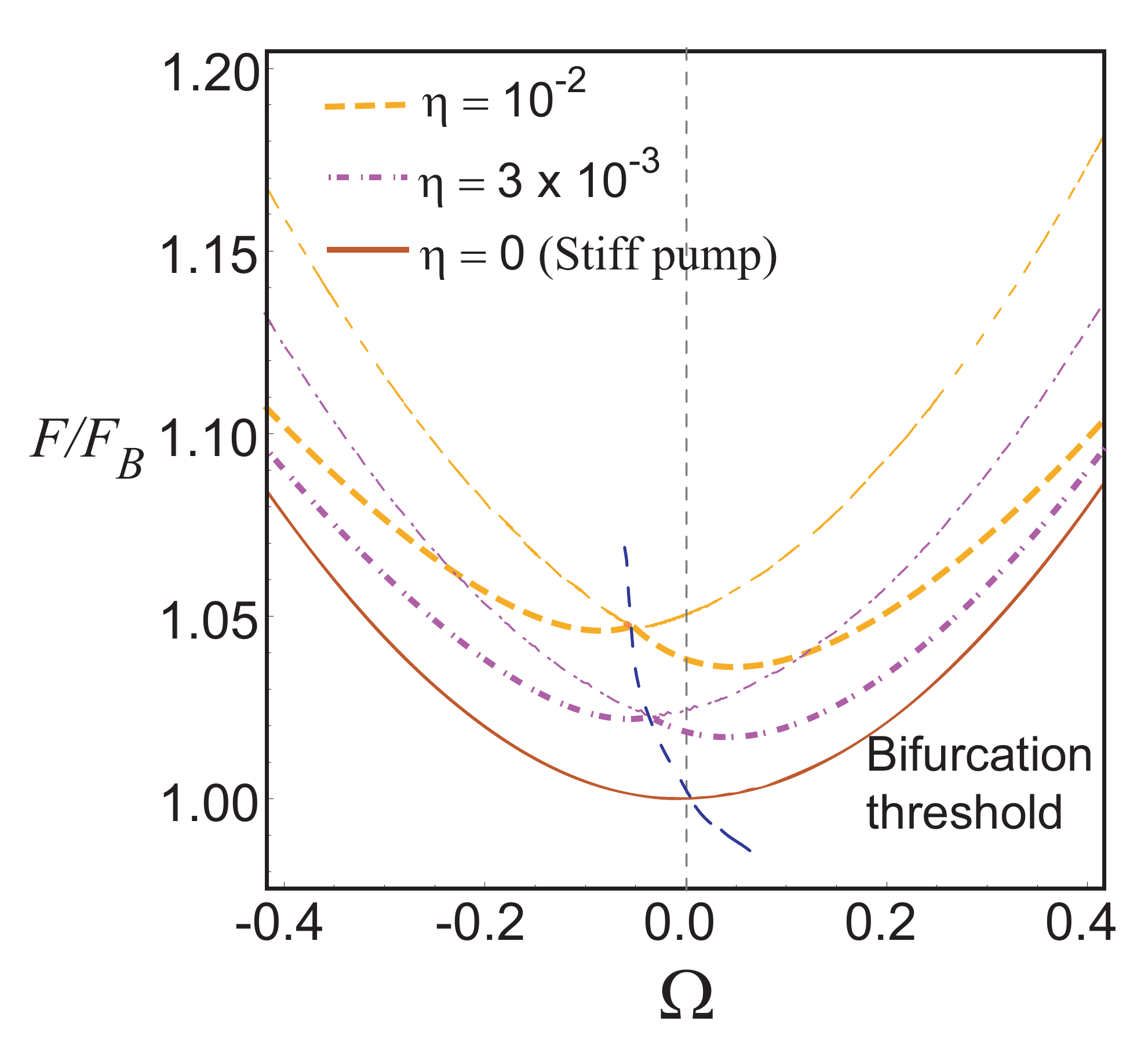}\\
\caption{Phase diagram for the DPA with back-action correction, for different values of effective temperature $\Theta_{\textrm{eff}}$. At each temperature, the curves corresponding to the two branches of noise back-action factor $\Pi$ are plotted parametrically (cf. Eq.\ref{main16}). The region between the two curves corresponds to the hysteresis of the system. The intersection of the two curves marks the locus of second-order phase transition points (dashed line). The bifurcation threshold is seen to shift upwards to higher pump strengths and towards higher values of $\Omega$ with temperature.}
\label{draw10}
\end{figure}

This can be further affirmed by looking at the phase diagram of the system. The locus of points showing the new bifurcation thresholds as a function of $\Omega$ determine the new phase boundary (Fig. \ref{draw10}). We note a shift of the phase boundary on incorporation of back-action which causes the minimum of the phase-transition curve to move away from $\Omega = 0$ (Fig. \ref{draw3}) towards negative frequency axis. This occurrence can be attributed to the decrease in resonance frequency $\omega_{0}$ with an increase in amplitude of oscillation due to higher drive strength (cf. Eq. (\ref{deltaeq}) for the typical form of dependence). This causes the effective detuning $\Omega = (\omega_{0} -\omega_{g})/\Gamma$ assume a non-zero negative value in the presence of noise. The above shift also explains the switch between the two branches of the bifurcation loci shown in Fig. \ref{draw10} for a given $\eta$. The system traces the upper (thin) curve showing a decrease in oscillation with drive, till the point it traverses through the minimum where it become unstable and switches to the other branch of high amplitude oscillation (thin portion of the other curve) where the resonance condition can be better met.
\par
Another important result highlighted by the phase diagram is the necessity of biasing the system near resonance. A comparison with the curves plotted in Fig. \ref{draw7} shows that the system is much more robust to back-action if biased near the relevant frequency (new $\omega_{0}$ corresponding to the shift in $\Omega$). For instance, $\Theta_{\textrm{eff}} = 0.1$ corresponded to a large shift in threshold power for bifurcation $\sim 20 \% $ (cf. Fig. \ref{draw7}c) while the corresponding curve in Fig. \ref{draw10} shows a shift of 5-10 \% in bifurcation threshold, if biased near new resonance.
%
%change \theta to \eta.
%
\par
It may be noted that for higher values of $\Omega (\geq 1)$, the frustration in the system increases further and the resultant corrections arising due to back-action approach order unity leading to anomalous behavior. Hence the underlying assumptions of perturbative treatment of back-action demands that we restrict ourselves to cases where the system is pumped sufficiently close to plasma frequency of the junction ($|\omega_{0} - \omega_{g}|\ll 1$).
%
%
%cccccccccccccccccccccccccccccccc
\section{Conclusions and Perspectives}
\label{summary}
%cccccccccccccccccccccccccccccccc
%
%
We performed a first-principle analysis of a Josephson parametric amplifier involving two RF energy sources and calculated the reciprocal effect of pump amplitude on signal and idler gains in a self-consistent manner. The analysis was based on a mean field approach of the intrinsic couplings between various components of the frequency spectrum, especially near bifurcation threshold. In particular, the effect of back-action on integrated noise power gain and self-oscillation amplitude response of the system were investigated.
\par
Starting with the equations derived in section \ref{model}, response functions for the DPA were calculated in section \ref{sec_stiff} for a stiff pump. Following analysis in section \ref{sec_pert_stiff} showed that the DPA can indeed achieve quantum limited noise temperature. The DPA also exhibited squeezing with fluctuations in one of the two quadratures squeezed below the quantum limit (cf. section \ref{sec_squ_stiff}). Maximum squeezing was found at the bifurcation threshold for a given pump detuning from resonance, as expected. Then in section \ref{sec_ba}, corrections to signal, idler and self-oscillation signal gains were calculated, in the presence of full back-action, yielding our main results as set of coupled equations - (\ref{force}), (\ref{main14}) and (\ref{main16})- involving three inter-dependent response functions of the system $\Xi$, $\Pi$ and signal, idler amplitdes.
\par
The significance of the inclusion of back-action is manifested even at the stiff pump level, as seen in section \ref{sec_st_stiff} for the calculation of self-oscillation amplitude.  It was shown that the inclusion of back-action to zeroth order (i.e. back-action due to self-oscillation signal only) is imperative to get steady state response of the system. These results are relevant for the amplifier designs based on three-wave mixing with SQUIDS as the inherent flux-based coupling mechanism impels a self-consistent treatment of pumps while calculating signal gains. We believe that the recent experimental results showing a spontaneous emission, reported for a flux-driven Josephson parametric amplifier \cite{yamamoto:042510}, can be well explained by doing a similar treatment. Another important feature that emerged out of this restricted analysis, with a stiff pump, was the coincidence of maximum-gain and bifurcation threshold. This harmony is contingent to symmetric bifurcation boundary obtained for the DPA.
\par
A self-consistent treatment of back-action led us to observe hysteresis in the system, as a function of drive strength. The magnitude of hysteresis increases as the detuning of the ghost frequency $\omega_{g}$ from resonance and effective temperature at the input port of the oscillator are increased. However, the feature of coincidence of maximum parametric gain and self-oscillation thresholds is preserved (as in the case of a stiff pump). An important conclusion of our work is the shift of the resonant frequency of the system itself. This can be clearly deduced from the shift of the phase boundary of the system, which shifts to higher drive strengths and higher resonant frequencies with increase in back-action. This indicates that adjusting the effective pump frequency $\omega_{g}$ near this `new' resonance is crucial for making the system less susceptible to back-action induced hysteresis and realize bifurcation at optimal pump strengths. Future experiments with devices employing the bifurcation of Josephson junction should take this effect into account. Also, the fact that DPA shows quantum limited behavior in the limit of large gains can be exploited to build a traveling wave amplifier using a cascaded chain of single stage amplifiers, with the effective noise temperature of each stage reduced to at least the cooling chamber of the next stage.
\par
To summarize, we have developed a minimal model inclusive of all the various components required to understand the dynamics of microwave parametric amplifiers based on purely dispersive elements like Josephson tunnel junctions, both away from and near the bifurcation threshold. The power and usefulness of this analysis lies in its generality which makes the techniques developed for the analysis of the DPA readily applicable to a host of systems operating either as three-wave or four-wave mixing devices.
%
%
%ccccccccccccccccccccccccccccccccccccccccc
\section*{Acknowledgements}

The authors wish to thank E. Akkermans, M. I. Dykman, S. M. Girvin, B. Huard, V. Manucharyan, R. J. Schoelkopf and A. D. Stone for useful discussions. A.K. is particularly indebted to R. Vijay for generously sharing all the details of his thesis work. The careful reading of the manuscript by Y. Nakamura and T. Yamamoto was very helpful for correcting several typos and is gratefully acknowledged. This work was supported by NSA through ARO Grant No. W911NF-05-01-0365, the Keck foundation, Agence Nationale pour la Recherche under the grant ANR07-CEXC-003 and the NSF through Grant No. DMR-032-5580. M.H.D. acknowledges partial support from College de France.
%ccccccccccccccccccccccccccccccccccccccccc
%
%
%ccccccccccccccccccc Michel's JBA Analysis cccccccccccccccccc
%
%
\appendix
\section{Phase Diagram for the Josephson Bifurcation Amplifier (JBA)}
\label{JBA_app}
\begin{figure}
\includegraphics[width=0.75\textwidth]{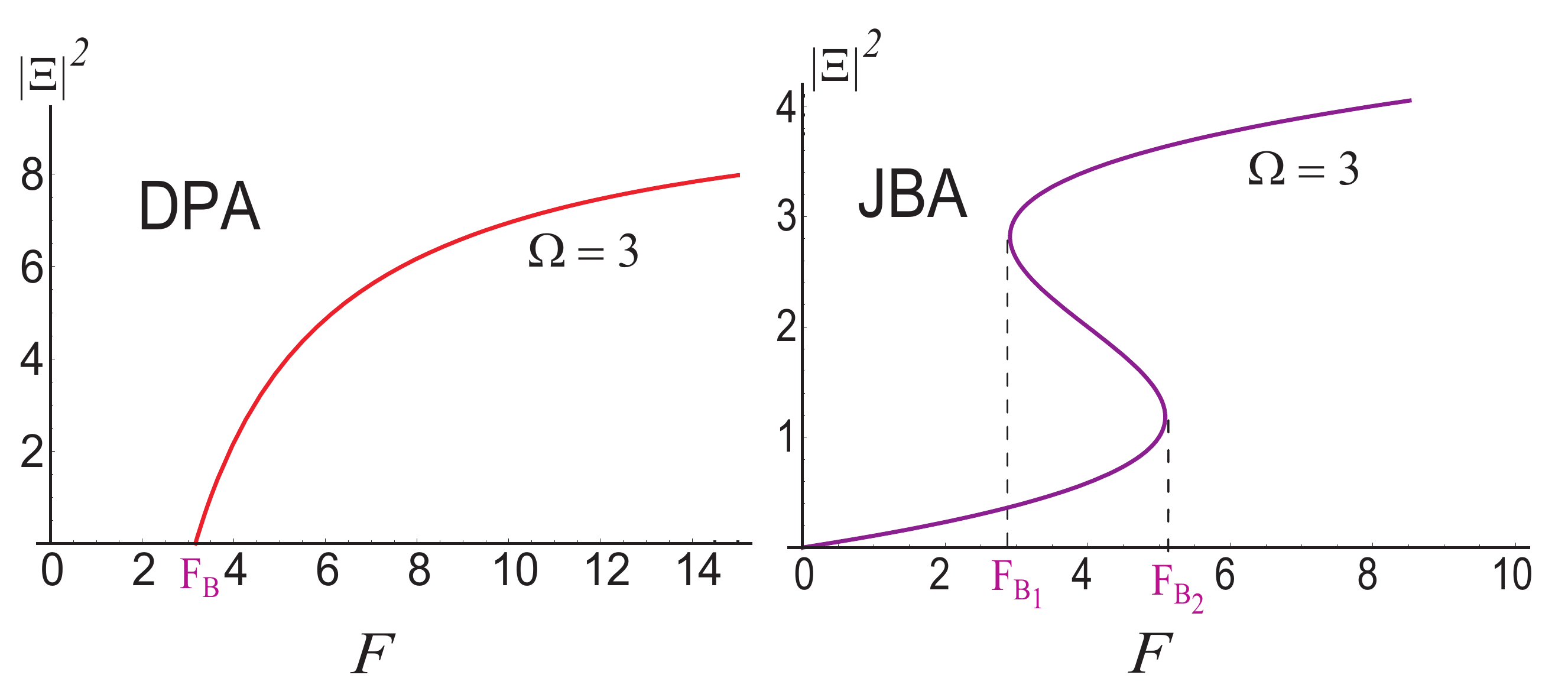}\\
\caption{Comparison between the steady state oscillation amplitude of the DPA (left panel) and the JBA (right panel), at the ghost and pump frequency respectively, in absence of noise. In both the case, these frequencies correspond to the average of the signal and the idler frequencies. Unlike the DPA, JBA displays a steady state oscillatory signal as soon as the pump is turned on. Note that the dimensionless pump power $F$ is related to the physical pump strength by different numerical factors in the two systems (see Eqns. \ref{DPAdrive} and \ref{JBAdrive}).}
\label{final}
\end{figure}

The equation of a damped RF-driven JTJ with a single pump, as in the model for the Josephson Bifurcation Amplifier (JBA), can be written as:
\begin{equation}
    \frac{d^{2}}{dt^{2}}\Phi + 2 \Gamma \frac{d}{dt}\Phi + \omega_{0}^2 \Phi
    \left[ 1-\lambda\left( \frac{2\pi \Phi }{\Phi_{0}}\right) ^{2}\right] =\frac{I_{RF}}{C}\cos \left( \omega t\right)
    \label{original}
\end{equation}
where $\lambda =1/6$, $\Gamma =\frac{1}{2RC}$ and $\omega_{0} =\frac{1}{C L_{J}}$.
As before, we assume a harmonic solution to the above equation
\begin{equation}
    \Xi\left( t\right) =\frac{1}{2}\left( \Xi\mathrm{e}^{-i\omega t} + \Xi^{\ast }\mathrm{e}^{+i\omega t}\right)
    \label{JBA_steady}
\end{equation}
which on performing harmonic balance and making the rotating wave approximation with $\delta (=\omega_{0} - \omega) ,\Gamma \ll \omega _{0}$, leads to
\begin{eqnarray*}
    \left[ -\omega _{0}^{2}+2\delta \omega _{0}+2i\omega _{0}\Gamma +\omega_{0}^{2}
    \left( 1- \chi \left\vert \Xi\right\vert ^{2}\right) \right] \Xi &=& f
\end{eqnarray*}
with $\chi = \frac{3 \lambda}{4}$ and $f =\omega_{0}^2 \frac{I_{RF}}{I_{0}}$. We can write the above equation as
\begin{eqnarray}
    \left[ \left( \delta -a\omega _{0}\left\vert \Xi\right\vert ^{2}\right)+i\Gamma \right] \Xi &=& \frac{f}{2\omega _{0}}
    \label{deltaeq}
\end{eqnarray}
where $a=\frac{\chi}{2} $ is the anharmonicity parameter (variation of resonant frequency with energy). Its value is $1/8E_{J}$ for the quadratic+quartic potential resulting from the expansion of the Josephson cosine potential
\begin{equation}
    \omega _{rp}=\omega _{0}\left( 1 - a\left\vert \Xi\right\vert ^{2}\right)
    + O\left[ \left\vert \Xi\right\vert ^{2}\right]
\end{equation}
Note that in the ``Transmon'' limit \cite{transmon}, this equation can be written as
\begin{equation*}
    \omega _{12}=\omega _{01}\left( 1-a \hbar \omega _{01}\right).
\end{equation*}
Using $\omega _{01}=\sqrt{8E_{J}E_{C}}/\hbar$, we recover the useful result
\begin{equation}
    \omega _{12}-\omega _{01}=E_{C}/\hbar.
\end{equation}
We can further reduce the equation for $\delta $ by introducing
\begin{eqnarray}
    \Omega = \frac{\delta }{\Gamma }, \; \;
    E = \frac{a\omega _{0}\left\vert \Xi\right\vert ^{2}}{\Gamma }, \;\;
    F = \frac{f^2}{4 \omega _{0}}\frac{a}{\Gamma ^{3}} =\frac{a}{4(\Gamma/\omega_{0})^3}\left(\frac{I_{RF}}{I_{0}}\right)^2
    \label{JBAdrive}
\end{eqnarray}
Using the above parameters, Eq. (\ref{deltaeq}) assumes the form
\begin{equation}
    \fbox{$\left[ \left(\Omega -E\right) ^{2}+1\right] E = F$}
    \label{maineq}
\end{equation}
For comparison, the corresponding state equation for the DPA is shown below (cf. Eq. \ref{stateeqDPA})
\begin{eqnarray}
    \boxed{\left[4(\Omega^2+1-F^2)+ 4\Omega F E + F^2 E^2\right]E = 0}.
\end{eqnarray}
where $E = \gamma |\Xi|^2$. A plot of the steady state behavior of the JBA is shown in Fig. \ref{final}. The analogous plot showing the steady state response of the DPA (as obtained in section \ref{sec_st_stiff}) is also displayed for comparison in the same figure. The plot for the DPA shows that before the bifurcation threshold, there is no response unlike the JBA. Also, note that there is no hysteresis in the DPA unlike the JBA, for which there is a lower and an upper bifurcation threshold. Thus the nature of transition between dynamical states at bifurcation is markedly different for the two systems.
For fixed $F$ the reduced detuning $\Omega$ can be obtained as a function of the reduced energy $E$
\begin{equation}
\Omega =E \pm \sqrt{\frac{F}{E}-1}
\end{equation}
Therefore (for real values of detuning $\Omega$), we demand $E_{\max}=F$. We can find the location of the bifurcation for a given F by requesting $\frac{d\Omega }{dE}=0$. Differentiating Eq. (\ref{maineq}), we get a condition for the extremum points.
\begin{equation}
    d\Omega \left[ 2E\left( \Omega -E\right) \right] + dE\left[ -2E\left( \Omega - E\right) +1+\left( \Omega -E\right) ^{2}\right] =0
\end{equation}
This shows that there exists a critical detuning and drive such that $E$ as a function of $B$ and $\Omega $ has a triple real root. This is the location of bifurcation in $E-\Omega$ plane. This can be found by locating the extremum of Eq. (\ref{maineq}) by using the above condition as a function of $E$ and requesting that they coincide. This gives us an equation for E
\begin{equation*}
    3E^{2}-4E\Omega +1+\Omega ^{2}=0
\end{equation*}
with two roots, whose values are given by:
\begin{eqnarray}
    E &=&\frac{2\Omega \pm \sqrt{4\Omega ^{2}-3\left( 1+\Omega ^{2}\right) }}{3} \nonumber\\
    &=&\frac{2\Omega \pm \sqrt{\Omega ^{2}-3}}{3}
\end{eqnarray}
They are degenerate for the critical values
\begin{eqnarray}
    \Omega _{c} = \sqrt{3}, \;\;\; E_{c} = \frac{2}{\sqrt{3}}, \;\;\; F_{c} = \frac{8}{3\sqrt{3}}.
\end{eqnarray}
In semiclassical terms, this critical point corresponds to the drive being such that the average energy leads to a detuning comparable to the line width.
\par
We reconsider Eq. (\ref{original}), but now include a weak `signal term' in addition to the pump
\begin{equation}
    \frac{d^{2}}{dt^{2}}X + 2\Gamma \frac{d}{dt}X + \omega _{0}^{2}X\left( 1-\chi X^{2}\right)
    = f\cos \left( \omega _{d}t\right) + \varepsilon \cos \left( \omega_{s}t\right)
\end{equation}
with $ \varepsilon \ll f$. We find a solution of the form
\begin{equation*}
    X\left( t\right) = \Xi\left( t\right) + \delta X \left( t\right)
\end{equation*}
where, $
    \delta X\left( t\right) = x\cos \left[ \left( \omega _{d}+\omega_{m}\right)t+\phi _{x}\right]
    + y\sin \left[ \left( \omega _{d}-\omega_{m}\right) t+\phi _{y}\right], \; \;
    \omega _{m} = \omega _{s}-\omega _{d}\ll \omega _{d} , \; \;
    \delta X\left( t\right) \ll \Xi\left( t\right).$
Under the above approximations, we can write the equation for $\delta X$ as
\begin{eqnarray}
    \frac{d^{2}}{dt^{2}}\delta X + 2\Gamma \frac{d}{dt}\delta X+\omega_{0}^{2}\delta X - 3\chi \omega _{0}^{2}\delta X
    \left[ \Xi\left(t\right)\right]^{2} & = & \varepsilon \cos \left( \omega _{s}t\right) \nonumber\\
    \frac{d^{2}}{dt^{2}}\delta X+2\Gamma \frac{d}{dt}\delta X+\omega_{0}^{2}\delta X - 3\chi \omega _{0}^{2}\delta X\left[ \frac{\left\vert
    \Xi\right\vert ^{2}}{2}+\frac{\Xi^{2}\mathrm{e}^{-2i\omega _{d}t} + c.c.}{4}\right]
    & = & \varepsilon \cos \left( \omega _{s}t\right)
\end{eqnarray}
where in the second step, we have used Eq. (\ref{JBA_steady}). Introducing
\begin{eqnarray*}
    \omega _{rs} &=&\omega _{0}\left( 1-\frac{3}{4}\chi \left\vert \Xi\right\vert ^{2}\right) \nonumber\\
    &=&\omega _{0}\left( 1 - 2a\left\vert \Xi\right\vert ^{2}\right) \neq \omega_{rp}
\end{eqnarray*}
we obtain
\begin{equation}
    \frac{d^{2}}{dt^{2}}\delta X+2\Gamma \frac{d}{dt}\delta X + \omega _{rs}^{2}%
    \left[ 1 + 2\left\vert \varepsilon \right\vert \sin \left( 2\omega_{d} + \varphi \right) \right] \delta X
    = \varepsilon \cos \left( \omega _{s}t\right).
\end{equation}
This is just the equation for a parametric amplifier. Therefore, the line $\Omega = 2E$ denotes the location of the optimal drive frequency for maximum parametric amplification. Note that it is not the same line as the zero-phase shift drive frequency ($\Omega =E$, the upper asymptote of the bistability line), nor the line passing through the location of the critical point ($\Omega =\frac{3}{2}E$), nor the lower asymptote of the bistability line ($\Omega =3E$). In the DPA, the optimal drive frequency for maximum parametric amplification remains $\Omega = 0$ below threshold.
\begin{figure}
\centering
\includegraphics[width=0.4\textwidth]{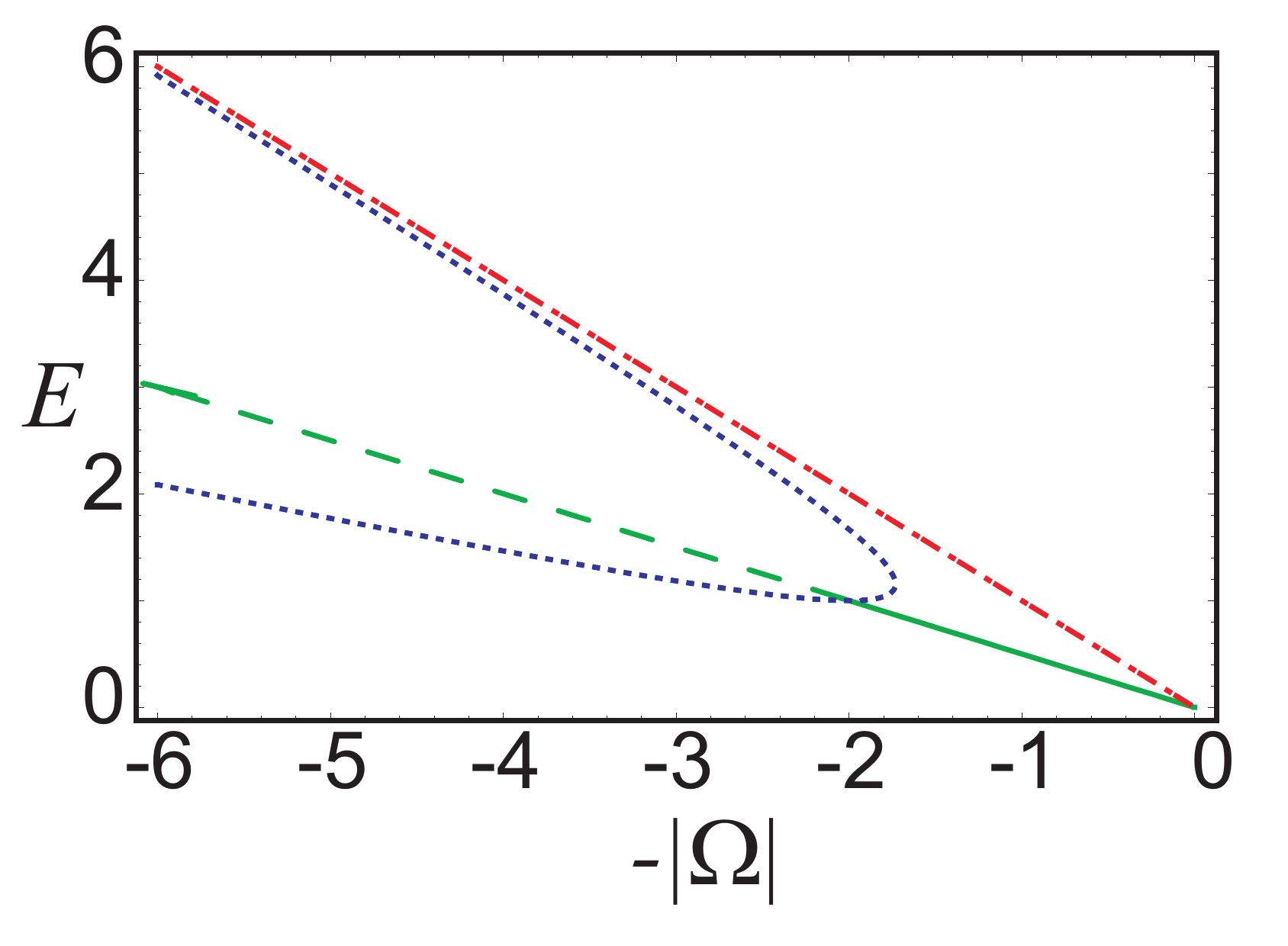}
\caption{Phase diagram of the JBA in E-$\Omega$ plane, showing the bifurcation curves (dotted-blue), the curve corresponding to the most efficient parametric amplification (solid-green with dashed portion indicating the bifurcation regime), and the asymptote to upper bifurcation level(dotted-dashed, red). Note that the $\Omega = 2 F$ line does \emph{not} intersect the bifurcation curve at the lowest value of detuning $\Omega = \sqrt{3}$ marking the commencement of bifurcation regime.}
\label{JBA_pd}
\end{figure}
The above bifurcation behavior is quite generic in nonlinear oscillators. The theory of these systems is fairly well developed and a detailed exposition of these ideas can be found in Ref. 43.
\section{Input-Output Theory}
\label{IOT_app}
Input-output theory (IOT) is a particular model of scattering theory (S-matrix theory) which applies to a system coupled to a heat bath. It is well documented in the literature but we include a brief description here for the help of the readers and consistency of notation. For the analysis via IOT, the resistance (R) of a circuit is replaced with a transmission line of characteristic impedance $Z_{c}$ (= R) and the voltage and the current along the line are expressed in terms of superposition of incoming and outgoing waves (Fig. \ref{IOmodel}). The waves represent either a signal launched on the line to drive the oscillator (pumps, signal) or the thermal/quantum fluctuations in the line (e.g. Nyquist noise of the resistor). The power of this semiclassical technique, apart from its calculational advantage, lies in the provision of simple physical insights into the link between the noise sources and dissipation. The voltages (V) and currents (I) are expressed in terms of incoming and outgoing field amplitudes (A) are expressed as:
\begin{figure}
\includegraphics[width=0.6\textwidth]{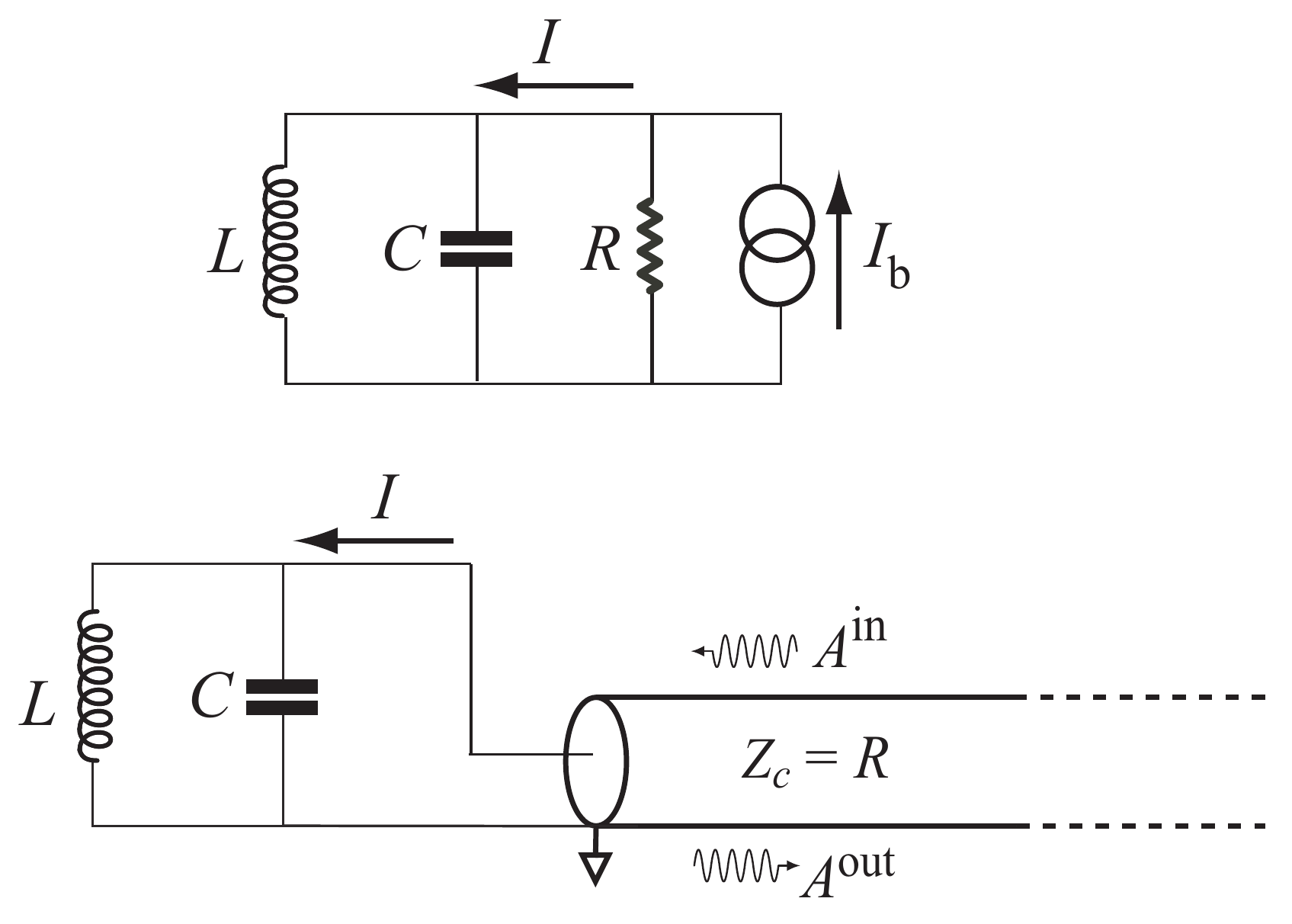}\\
\caption{Top panel shows a damped LC oscillator driven by an RF current source. Bottom panel shows the equivalent circuit in which the current source and its internal resistance have been replaced by a semi-infinite transmission line (input output analog). The two cases are equivalent from the point of view of the LC oscillator if we make the identification $Z_{C} = R$ and $A^{\textrm{in}} = \sqrt{R} I_{\textrm{b}}/2$.}
\label{IOmodel}
\end{figure}
\begin{eqnarray}
     V(z,t) = \sqrt{Z_{c}} \left( A^{\textrm{out}}(z,t) + A^{\textrm{in}}(z,t)\right)\nonumber\\
     I(z,t) = \frac{1}{\sqrt{Z_{c}}} \left( A^{\textrm{out}}(z,t) - A^{\textrm{in}}(z,t)\right)
\end{eqnarray}
It is straightforward to obtain constitutive relations linking input, output and internal fields of the amplifier by imposing the appropriate boundary conditions at the termination of the line (z=0)
\begin{eqnarray}
     V(t) = V^{\textrm{in}}(t) + V^{\textrm{out}}(t); \; \;\;\;\; I(t) = I^{\textrm{in}}(t) - I^{\textrm{out}}(t)
\end{eqnarray}
where we have used the relation $V^{\textrm{in/out}} = \sqrt{Z_{c}} A^{\textrm{in/out}}$ and $I^{\textrm{in/out}} = \frac{A^{\textrm{in/out}}}{\sqrt{Z_{c}}}$.
\par
It is useful to define the quantities $a[\omega]$ as
\begin{eqnarray}
  \sqrt{\frac{\hbar\omega}{2}}a[\omega] = A[\omega]
  \label{ampcomm}
\end{eqnarray}
where $A[\omega] = \frac{1}{\sqrt{2 \pi}}\int dt A(t) \exp(\dot{\imath}\omega t)$.
This leads us to a natural generalization to quantum regime as the normalized field amplitude $a$ plays the role of bosonic field operator as defined for a harmonic oscillator. It obeys the following well-known field theoretical commutation relation \cite{Reynaud_qmfluc}
\begin{eqnarray}
    [\hat{a}^{\textrm{in}}[\omega],\hat{a}^{\textrm{in}}[\omega']] = \textrm{sgn}[\omega]\;\; \delta (\omega+\omega')
    \label{comm}
\end{eqnarray}
The fluctuations of field creation and annihilation operators is characterized by the corresponding noise spectrum in thermal equilibrium
\begin{eqnarray}
    \langle \left\{\hat{a}^{\textrm{in}}[\omega],\; \hat{a}^{\textrm{in}}[\omega']\right\}\rangle_{T} & = &  \;\;S_{aa}[\omega]\;\; \delta (\omega+\omega')\nonumber\\
    S_{aa}[\omega] & = & \left[\frac{1}{\exp(\frac{\hbar\omega}{k_{B} T}) - 1} + \frac{1}{2} \right]
    =\frac{1}{2} \coth{\frac{\hbar |\omega|}{2 k_{B} T}}.
    \label{anticomm}
\end{eqnarray}
The quantity $\hbar\omega S_{aa}^{\textrm{in}}$ denotes the total energy per mode and reduces to $\frac{\hbar\omega}{2}$ in the limit of zero temperature (vacuum fluctuations) and the classical limit of $k_{B} T$ in the limit of high temperature. Eqns. (\ref{comm}) and (\ref{anticomm}) are valid over the entire frequency range, including the negative frequencies. We can return to the conventional description restricted to only positive frequencies by the identification
\begin{center}
    $\hat{a}[-\omega] \rightarrow \hat{a}^{\dag}[\omega]$.
\end{center}
The preceding equations lead us to define the ordered spectral density
\begin{eqnarray}
    \langle \hat{A}[\omega]\hat{A}[\omega']\rangle  = \frac{\hbar\omega}{4}\left[\coth\left(\frac{\hbar\omega}{2 k_{B} T}\right) + 1 \right]\delta(\omega+\omega').
\end{eqnarray}
Thus, we can easily write the fluctuations of the voltage across the resistor
\begin{eqnarray}
    \langle \hat{V}[\omega]\hat{V}[\omega']\rangle
    & = & \frac{Z_{c}\hbar\omega}{4}\left[\coth\left(\frac{\hbar\omega}{2 k_{B} T}\right) + 1 \right]\delta(\omega+\omega')
    \label{voltfluc}
\end{eqnarray}
which follows from $V^{\textrm{in/out}} = \sqrt{Z_{c}} A^{\textrm{in/out}}$. Eq. (\ref{voltfluc}) is used in section \ref{corr_calc} while calculating the noise back-action factor $\Pi$.
\par
The validity of this crossover to quantum description lies in the fact that in case of parametric interaction, the difference between the classical and quantum evolution vanishes when the number of photons in the line is large or the coupling of the system to reservoir is weak \cite{PhysRevA.46.2766}. We can then regard the quantum fluctuations to be driven by classical random fields, obeying classical equations of motion.
\section{}
\label{Symbols_app}
\begin{center}
\begin{tabular}{|c|l|}
\hline
\multicolumn{2}{|c|}{Table of Symbols} \\
\hline
$\Delta$ & Reduced detuning of signal from $\omega_{g}$ $\left(=\frac{\omega_{S} -\omega_{g}}{\Gamma}\right)$ \\
$E_{J}$ & Josephson energy\\
$F$ & Effective pump strength\\
$F_{BA}$ & Back-action corrected pump strength\\
$G$ & Power gain of the amplifier\\
$\Gamma$ & Damping rate $(\frac{1}{2 R C})$ \\
$\gamma $ & Effective nonlinearity parameter $\left(=\frac{2\lambda}{(1 +\frac{\omega_{1}}{\omega_{0}})(1 +\frac{\omega_{2}}{\omega_{0}})}\right)$\\
$\Theta_{\textrm{eff}} $ & Effective temperature at the input port $(= \frac{k_{B} T}{E_{J}})$\\
$\eta $ & Back-action index $ (=\gamma \cdot\Theta_{\textrm{eff}})$\\
$I_{N}$ & Noise current \\
$I_{RF}$ & RF drive current \\
$L_{J}$ & Josephson inductance \\
$\Phi$ & Flux across the Josephson Junction\\
$\Phi_{0}$ & Flux quantum $(=\frac{h}{2e})$\\
$\varphi$ & Dimensionless flux variable\\
$\Pi$ & Noise back-action factor\\
$\omega_{0}$ & Resonant frequency of Josephson oscillator \\
$\omega_{1}$ & Pump frequency 1 \\
$\omega_{1}$ & Pump frequency 2\\
$\omega_{g}$ & Ghost frequency $\left( = \frac{\omega_{1} + \omega_{2}}{2} \right)$ \\
$\Omega$ & Reduced detuning of $\omega_{g}$ from $\omega_{0}$ $\left(=\frac{\omega_{0} -\omega_{g}}{\Gamma}\right)$\\
$r$ & Reflection coefficient \\
$s$ & Interconversion gain coefficient \\
$S_{aa}$ & Photon number spectral density \\
$T$ & Black body temperature at the input (signal and idler ports)\\
$T_{N}$ & Noise temperature of the amplifier \\
$X[\omega]$ & Amplitude of noise component at generic frequency $\omega$\\
$\Xi$ & Steady state amplitude phasor\\
$Y$ & Amplitude phasor for pump 1\\
$Z$ & Amplitude phasor for pump 2\\ \hline
\end{tabular}
\end{center}
\linespread{2.}
\bibliographystyle{apsrev}

\begin{thebibliography}{0}
\expandafter\ifx\csname natexlab\endcsname\relax\def\natexlab#1{#1}\fi
\expandafter\ifx\csname bibnamefont\endcsname\relax
  \def\bibnamefont#1{#1}\fi
\expandafter\ifx\csname bibfnamefont\endcsname\relax
  \def\bibfnamefont#1{#1}\fi
\expandafter\ifx\csname citenamefont\endcsname\relax
  \def\citenamefont#1{#1}\fi
\expandafter\ifx\csname url\endcsname\relax
  \def\url#1{\texttt{#1}}\fi
\expandafter\ifx\csname urlprefix\endcsname\relax\def\urlprefix{URL }\fi
\providecommand{\bibinfo}[2]{#2}
\providecommand{\eprint}[2][]{\url{#2}}

\end{thebibliography}


\begin{thebibliography}{45}
\expandafter\ifx\csname natexlab\endcsname\relax\def\natexlab#1{#1}\fi
\expandafter\ifx\csname bibnamefont\endcsname\relax
  \def\bibnamefont#1{#1}\fi
\expandafter\ifx\csname bibfnamefont\endcsname\relax
  \def\bibfnamefont#1{#1}\fi
\expandafter\ifx\csname citenamefont\endcsname\relax
  \def\citenamefont#1{#1}\fi
\expandafter\ifx\csname url\endcsname\relax
  \def\url#1{\texttt{#1}}\fi
\expandafter\ifx\csname urlprefix\endcsname\relax\def\urlprefix{URL }\fi
\providecommand{\bibinfo}[2]{#2}
\providecommand{\eprint}[2][]{\url{#2}}

\bibitem[{\citenamefont{Knobel and Cleland}(2003)}]{NEMS1}
\bibinfo{author}{\bibfnamefont{R.~G.} \bibnamefont{Knobel}} \bibnamefont{and}
  \bibinfo{author}{\bibfnamefont{A.~N.} \bibnamefont{Cleland}},
  \bibinfo{journal}{Nature} \textbf{\bibinfo{volume}{304}},
  \bibinfo{pages}{291} (\bibinfo{year}{2003}).

\bibitem[{\citenamefont{LaHaye et~al.}(2004)\citenamefont{LaHaye, Buu,
  Camarota, and Schwab}}]{NEMS2}
\bibinfo{author}{\bibfnamefont{M.~D.} \bibnamefont{LaHaye}},
  \bibinfo{author}{\bibfnamefont{O.}~\bibnamefont{Buu}},
  \bibinfo{author}{\bibfnamefont{B.}~\bibnamefont{Camarota}}, \bibnamefont{and}
  \bibinfo{author}{\bibfnamefont{K.~C.} \bibnamefont{Schwab}},
  \bibinfo{journal}{Science} \textbf{\bibinfo{volume}{304}},
  \bibinfo{pages}{74} (\bibinfo{year}{2004}).

\bibitem[{\citenamefont{Yang et~al.}(2006)\citenamefont{Yang, Callegari, Feng,
  Ekinci, and Roukes}}]{NEMS3}
\bibinfo{author}{\bibfnamefont{Y.~T.} \bibnamefont{Yang}},
  \bibinfo{author}{\bibfnamefont{C.}~\bibnamefont{Callegari}},
  \bibinfo{author}{\bibfnamefont{X.~L.} \bibnamefont{Feng}},
  \bibinfo{author}{\bibfnamefont{K.~L.} \bibnamefont{Ekinci}},
  \bibnamefont{and} \bibinfo{author}{\bibfnamefont{M.~L.}
  \bibnamefont{Roukes}}, \bibinfo{journal}{Nano Letters}
  \textbf{\bibinfo{volume}{6}}, \bibinfo{pages}{583} (\bibinfo{year}{2006}).

\bibitem[{\citenamefont{Duan et~al.}(2000)\citenamefont{Duan, Cirac, Zoller,
  and Polzik}}]{OPT1}
\bibinfo{author}{\bibfnamefont{L.-M.} \bibnamefont{Duan}},
  \bibinfo{author}{\bibfnamefont{J.~I.} \bibnamefont{Cirac}},
  \bibinfo{author}{\bibfnamefont{P.}~\bibnamefont{Zoller}}, \bibnamefont{and}
  \bibinfo{author}{\bibfnamefont{E.~S.} \bibnamefont{Polzik}},
  \bibinfo{journal}{Phys. Rev. Lett.} \textbf{\bibinfo{volume}{85}},
  \bibinfo{pages}{5643} (\bibinfo{year}{2000}).

\bibitem[{\citenamefont{Duan et~al.}(2001)\citenamefont{Duan, Lukin, Cirac, and
  Zoller}}]{OPT2}
\bibinfo{author}{\bibfnamefont{L.-M.} \bibnamefont{Duan}},
  \bibinfo{author}{\bibfnamefont{M.~D.} \bibnamefont{Lukin}},
  \bibinfo{author}{\bibfnamefont{J.~I.} \bibnamefont{Cirac}}, \bibnamefont{and}
  \bibinfo{author}{\bibfnamefont{P.}~\bibnamefont{Zoller}},
  \bibinfo{journal}{Nature} \textbf{\bibinfo{volume}{414}},
  \bibinfo{pages}{413} (\bibinfo{year}{2001}).

\bibitem[{\citenamefont{Caves}(1982)}]{PhysRevD.26.1817}
\bibinfo{author}{\bibfnamefont{C.~M.} \bibnamefont{Caves}},
  \bibinfo{journal}{Phys. Rev. D} \textbf{\bibinfo{volume}{26}},
  \bibinfo{pages}{1817} (\bibinfo{year}{1982}).

\bibitem[{\citenamefont{Stolen and Bjorkholm}(1982)}]{1071660}
\bibinfo{author}{\bibfnamefont{R.}~\bibnamefont{Stolen}} \bibnamefont{and}
  \bibinfo{author}{\bibfnamefont{J.}~\bibnamefont{Bjorkholm}},
  \bibinfo{journal}{IEEE J. Quantum Electron.} \textbf{\bibinfo{volume}{18}},
  \bibinfo{pages}{1062} (\bibinfo{year}{1982}), ISSN \bibinfo{issn}{0018-9197}.

\bibitem[{\citenamefont{Caves and Crouch}(1987)}]{Caves:87}
\bibinfo{author}{\bibfnamefont{C.~M.} \bibnamefont{Caves}} \bibnamefont{and}
  \bibinfo{author}{\bibfnamefont{D.~D.} \bibnamefont{Crouch}},
  \bibinfo{journal}{J. Opt. Soc. Am. B} \textbf{\bibinfo{volume}{4}},
  \bibinfo{pages}{1535} (\bibinfo{year}{1987}).

\bibitem[{\citenamefont{Grangier et~al.}(1998)\citenamefont{Grangier, Levenson,
  and Poizat}}]{opticsparamp}
\bibinfo{author}{\bibfnamefont{P.}~\bibnamefont{Grangier}},
  \bibinfo{author}{\bibfnamefont{J.~A.} \bibnamefont{Levenson}},
  \bibnamefont{and} \bibinfo{author}{\bibfnamefont{J.-P.}
  \bibnamefont{Poizat}}, \bibinfo{journal}{Nature}
  \textbf{\bibinfo{volume}{396}}, \bibinfo{pages}{537} (\bibinfo{year}{1998}).

\bibitem[{\citenamefont{Hansryd et~al.}(2002)\citenamefont{Hansryd, Andrekson,
  Westlund, Li, and Hedekvist}}]{1016354}
\bibinfo{author}{\bibfnamefont{J.}~\bibnamefont{Hansryd}},
  \bibinfo{author}{\bibfnamefont{P.}~\bibnamefont{Andrekson}},
  \bibinfo{author}{\bibfnamefont{M.}~\bibnamefont{Westlund}},
  \bibinfo{author}{\bibfnamefont{J.}~\bibnamefont{Li}}, \bibnamefont{and}
  \bibinfo{author}{\bibfnamefont{P.-O.} \bibnamefont{Hedekvist}},
  \bibinfo{journal}{IEEE Journal of Selected Topics in Quantum Electronics}
  \textbf{\bibinfo{volume}{8}}, \bibinfo{pages}{506} (\bibinfo{year}{2002}),
  ISSN \bibinfo{issn}{1077-260X}.

\bibitem[{\citenamefont{Radic et~al.}(2003)\citenamefont{Radic, McKinstrie,
  Jopson, Centanni, Lin, and Agrawal}}]{1207224}
\bibinfo{author}{\bibfnamefont{S.}~\bibnamefont{Radic}},
  \bibinfo{author}{\bibfnamefont{C.}~\bibnamefont{McKinstrie}},
  \bibinfo{author}{\bibfnamefont{R.}~\bibnamefont{Jopson}},
  \bibinfo{author}{\bibfnamefont{J.}~\bibnamefont{Centanni}},
  \bibinfo{author}{\bibfnamefont{Q.}~\bibnamefont{Lin}}, \bibnamefont{and}
  \bibinfo{author}{\bibfnamefont{G.}~\bibnamefont{Agrawal}},
  \bibinfo{journal}{Electron. Lett.} \textbf{\bibinfo{volume}{39}},
  \bibinfo{pages}{838} (\bibinfo{year}{2003}), ISSN \bibinfo{issn}{0013-5194}.

\bibitem[{\citenamefont{Wadefalk et~al.}(2003)\citenamefont{Wadefalk, Mellberg,
  Angelov, Barsky, Bui, Choumas, Grundbacher, Kollberg, Lai, Rorsman
  et~al.}}]{1201803}
\bibinfo{author}{\bibfnamefont{N.}~\bibnamefont{Wadefalk}},
  \bibinfo{author}{\bibfnamefont{A.}~\bibnamefont{Mellberg}},
  \bibinfo{author}{\bibfnamefont{I.}~\bibnamefont{Angelov}},
  \bibinfo{author}{\bibfnamefont{M.}~\bibnamefont{Barsky}},
  \bibinfo{author}{\bibfnamefont{S.}~\bibnamefont{Bui}},
  \bibinfo{author}{\bibfnamefont{E.}~\bibnamefont{Choumas}},
  \bibinfo{author}{\bibfnamefont{R.}~\bibnamefont{Grundbacher}},
  \bibinfo{author}{\bibfnamefont{E.}~\bibnamefont{Kollberg}},
  \bibinfo{author}{\bibfnamefont{R.}~\bibnamefont{Lai}},
  \bibinfo{author}{\bibfnamefont{N.}~\bibnamefont{Rorsman}},
  \bibnamefont{et~al.}, \bibinfo{journal}{IEEE Trans. Microwave Theory Tech.}
  \textbf{\bibinfo{volume}{51}}, \bibinfo{pages}{1705} (\bibinfo{year}{2003}),
  ISSN \bibinfo{issn}{0018-9480}.

\bibitem[{\citenamefont{Bradley et~al.}(2003)\citenamefont{Bradley, Clarke,
  Kinion, Rosenberg, van Bibber, Matsuki, M\"uck, and
  Sikivie}}]{RevModPhys.75.777}
\bibinfo{author}{\bibfnamefont{R.}~\bibnamefont{Bradley}},
  \bibinfo{author}{\bibfnamefont{J.}~\bibnamefont{Clarke}},
  \bibinfo{author}{\bibfnamefont{D.}~\bibnamefont{Kinion}},
  \bibinfo{author}{\bibfnamefont{L.~J.} \bibnamefont{Rosenberg}},
  \bibinfo{author}{\bibfnamefont{K.}~\bibnamefont{van Bibber}},
  \bibinfo{author}{\bibfnamefont{S.}~\bibnamefont{Matsuki}},
  \bibinfo{author}{\bibfnamefont{M.}~\bibnamefont{M\"uck}}, \bibnamefont{and}
  \bibinfo{author}{\bibfnamefont{P.}~\bibnamefont{Sikivie}},
  \bibinfo{journal}{Rev. Mod. Phys.} \textbf{\bibinfo{volume}{75}},
  \bibinfo{pages}{777} (\bibinfo{year}{2003}).

\bibitem[{\citenamefont{Wallraff et~al.}(2004)\citenamefont{Wallraff, Schuster,
  Blais, Frunzio, Huang, Majer, Kumar, Girvin, and Schoelkopf}}]{RSL1}
\bibinfo{author}{\bibfnamefont{A.}~\bibnamefont{Wallraff}},
  \bibinfo{author}{\bibfnamefont{D.~I.} \bibnamefont{Schuster}},
  \bibinfo{author}{\bibfnamefont{A.}~\bibnamefont{Blais}},
  \bibinfo{author}{\bibfnamefont{L.}~\bibnamefont{Frunzio}},
  \bibinfo{author}{\bibfnamefont{R.-S.} \bibnamefont{Huang}},
  \bibinfo{author}{\bibfnamefont{J.}~\bibnamefont{Majer}},
  \bibinfo{author}{\bibfnamefont{S.}~\bibnamefont{Kumar}},
  \bibinfo{author}{\bibfnamefont{S.~M.} \bibnamefont{Girvin}},
  \bibnamefont{and} \bibinfo{author}{\bibfnamefont{R.~J.}
  \bibnamefont{Schoelkopf}}, \bibinfo{journal}{Nature}
  \textbf{\bibinfo{volume}{431}}, \bibinfo{pages}{162} (\bibinfo{year}{2004}).

\bibitem[{\citenamefont{Movshovich et~al.}(1991)\citenamefont{Movshovich,
  Yurke, Smith, and Silver}}]{PhysRevLett.67.1411}
\bibinfo{author}{\bibfnamefont{R.}~\bibnamefont{Movshovich}},
  \bibinfo{author}{\bibfnamefont{B.}~\bibnamefont{Yurke}},
  \bibinfo{author}{\bibfnamefont{A.~D.} \bibnamefont{Smith}}, \bibnamefont{and}
  \bibinfo{author}{\bibfnamefont{A.~H.} \bibnamefont{Silver}},
  \bibinfo{journal}{Phys. Rev. Lett.} \textbf{\bibinfo{volume}{67}},
  \bibinfo{pages}{1411} (\bibinfo{year}{1991}).

\bibitem[{\citenamefont{Siddiqi et~al.}(2004)\citenamefont{Siddiqi, Vijay,
  Pierre, Wilson, Metcalfe, Rigetti, Frunzio, and
  Devoret}}]{PhysRevLett.93.207002}
\bibinfo{author}{\bibfnamefont{I.}~\bibnamefont{Siddiqi}},
  \bibinfo{author}{\bibfnamefont{R.}~\bibnamefont{Vijay}},
  \bibinfo{author}{\bibfnamefont{F.}~\bibnamefont{Pierre}},
  \bibinfo{author}{\bibfnamefont{C.~M.} \bibnamefont{Wilson}},
  \bibinfo{author}{\bibfnamefont{M.}~\bibnamefont{Metcalfe}},
  \bibinfo{author}{\bibfnamefont{C.}~\bibnamefont{Rigetti}},
  \bibinfo{author}{\bibfnamefont{L.}~\bibnamefont{Frunzio}}, \bibnamefont{and}
  \bibinfo{author}{\bibfnamefont{M.~H.} \bibnamefont{Devoret}},
  \bibinfo{journal}{Phys. Rev. Lett.} \textbf{\bibinfo{volume}{93}},
  \bibinfo{pages}{207002} (\bibinfo{year}{2004}).

\bibitem[{\citenamefont{Metcalfe
  et~al.}(2007{\natexlab{a}})\citenamefont{Metcalfe, Boaknin, Manucharyan,
  Vijay, Siddiqi, Rigetti, Frunzio, Schoelkopf, and Devoret}}]{metcalfe:174516}
\bibinfo{author}{\bibfnamefont{M.}~\bibnamefont{Metcalfe}},
  \bibinfo{author}{\bibfnamefont{E.}~\bibnamefont{Boaknin}},
  \bibinfo{author}{\bibfnamefont{V.}~\bibnamefont{Manucharyan}},
  \bibinfo{author}{\bibfnamefont{R.}~\bibnamefont{Vijay}},
  \bibinfo{author}{\bibfnamefont{I.}~\bibnamefont{Siddiqi}},
  \bibinfo{author}{\bibfnamefont{C.}~\bibnamefont{Rigetti}},
  \bibinfo{author}{\bibfnamefont{L.}~\bibnamefont{Frunzio}},
  \bibinfo{author}{\bibfnamefont{R.~J.} \bibnamefont{Schoelkopf}},
  \bibnamefont{and} \bibinfo{author}{\bibfnamefont{M.~H.}
  \bibnamefont{Devoret}}, \bibinfo{journal}{Phy. Rev. B}
  \textbf{\bibinfo{volume}{76}}, \bibinfo{eid}{174516}
  (\bibinfo{year}{2007}{\natexlab{a}}).

\bibitem[{\citenamefont{Tholen et~al.}(2007)\citenamefont{Tholen, Ergul,
  Doherty, Weber, Gregis, and Haviland}}]{tholen:253509}
\bibinfo{author}{\bibfnamefont{E.~A.} \bibnamefont{Tholen}},
  \bibinfo{author}{\bibfnamefont{A.}~\bibnamefont{Ergul}},
  \bibinfo{author}{\bibfnamefont{E.~M.} \bibnamefont{Doherty}},
  \bibinfo{author}{\bibfnamefont{F.~M.} \bibnamefont{Weber}},
  \bibinfo{author}{\bibfnamefont{F.}~\bibnamefont{Gregis}}, \bibnamefont{and}
  \bibinfo{author}{\bibfnamefont{D.~B.} \bibnamefont{Haviland}},
  \bibinfo{journal}{Appl. Phys. Lett.} \textbf{\bibinfo{volume}{90}},
  \bibinfo{pages}{253509} (\bibinfo{year}{2007}).

\bibitem[{\citenamefont{Castellanos-Beltran and
  Lehnert}(2007)}]{castellanos-beltran:083509}
\bibinfo{author}{\bibfnamefont{M.~A.} \bibnamefont{Castellanos-Beltran}}
  \bibnamefont{and} \bibinfo{author}{\bibfnamefont{K.~W.}
  \bibnamefont{Lehnert}}, \bibinfo{journal}{Appl. Phys. Lett.}
  \textbf{\bibinfo{volume}{91}}, \bibinfo{pages}{083509}
  (\bibinfo{year}{2007}).

\bibitem[{\citenamefont{Yamamoto et~al.}(2008)\citenamefont{Yamamoto, Inomata,
  Watanabe, Matsuba, Miyazaki, Oliver, Nakamura, and Tsai}}]{yamamoto:042510}
\bibinfo{author}{\bibfnamefont{T.}~\bibnamefont{Yamamoto}},
  \bibinfo{author}{\bibfnamefont{K.}~\bibnamefont{Inomata}},
  \bibinfo{author}{\bibfnamefont{M.}~\bibnamefont{Watanabe}},
  \bibinfo{author}{\bibfnamefont{K.}~\bibnamefont{Matsuba}},
  \bibinfo{author}{\bibfnamefont{T.}~\bibnamefont{Miyazaki}},
  \bibinfo{author}{\bibfnamefont{W.~D.} \bibnamefont{Oliver}},
  \bibinfo{author}{\bibfnamefont{Y.}~\bibnamefont{Nakamura}}, \bibnamefont{and}
  \bibinfo{author}{\bibfnamefont{J.~S.} \bibnamefont{Tsai}},
  \bibinfo{journal}{Appl. Phys. Lett.} \textbf{\bibinfo{volume}{93}},
  \bibinfo{pages}{042510} (\bibinfo{year}{2008}).

\bibitem[{\citenamefont{Boulant et~al.}(2007)\citenamefont{Boulant, Ithier,
  Meeson, Nguyen, Vion, Esteve, Siddiqi, Vijay, Rigetti, Pierre
  et~al.}}]{boulant:014525}
\bibinfo{author}{\bibfnamefont{N.}~\bibnamefont{Boulant}},
  \bibinfo{author}{\bibfnamefont{G.}~\bibnamefont{Ithier}},
  \bibinfo{author}{\bibfnamefont{P.}~\bibnamefont{Meeson}},
  \bibinfo{author}{\bibfnamefont{F.}~\bibnamefont{Nguyen}},
  \bibinfo{author}{\bibfnamefont{D.}~\bibnamefont{Vion}},
  \bibinfo{author}{\bibfnamefont{D.}~\bibnamefont{Esteve}},
  \bibinfo{author}{\bibfnamefont{I.}~\bibnamefont{Siddiqi}},
  \bibinfo{author}{\bibfnamefont{R.}~\bibnamefont{Vijay}},
  \bibinfo{author}{\bibfnamefont{C.}~\bibnamefont{Rigetti}},
  \bibinfo{author}{\bibfnamefont{F.}~\bibnamefont{Pierre}},
  \bibnamefont{et~al.}, \bibinfo{journal}{Phys. Rev. B}
  \textbf{\bibinfo{volume}{76}}, \bibinfo{eid}{014525} (\bibinfo{year}{2007}).

\bibitem[{\citenamefont{Yariv}(1997)}]{Yariv}
\bibinfo{author}{\bibfnamefont{A.}~\bibnamefont{Yariv}},
  \emph{\bibinfo{title}{Optical Electronics in Modern Communications}}
  (\bibinfo{publisher}{Oxford University Press}, \bibinfo{year}{1997}).

\bibitem[{\citenamefont{Boggio et~al.}(2001)\citenamefont{Boggio, Tenenbaum,
  and Fragnito}}]{Boggio:01}
\bibinfo{author}{\bibfnamefont{J.~M.~C.} \bibnamefont{Boggio}},
  \bibinfo{author}{\bibfnamefont{S.}~\bibnamefont{Tenenbaum}},
  \bibnamefont{and} \bibinfo{author}{\bibfnamefont{H.~L.}
  \bibnamefont{Fragnito}}, \bibinfo{journal}{J. Opt. Soc. Am. B}
  \textbf{\bibinfo{volume}{18}}, \bibinfo{pages}{1428} (\bibinfo{year}{2001}).

\bibitem[{\citenamefont{Bogris et~al.}(2005)\citenamefont{Bogris, Syvridis,
  Kylemark, and Andrekson}}]{1506859}
\bibinfo{author}{\bibfnamefont{A.}~\bibnamefont{Bogris}},
  \bibinfo{author}{\bibfnamefont{D.}~\bibnamefont{Syvridis}},
  \bibinfo{author}{\bibfnamefont{P.}~\bibnamefont{Kylemark}}, \bibnamefont{and}
  \bibinfo{author}{\bibfnamefont{P.}~\bibnamefont{Andrekson}},
  \bibinfo{journal}{J. Lightwave Tech.} \textbf{\bibinfo{volume}{23}},
  \bibinfo{pages}{2788} (\bibinfo{year}{2005}), ISSN \bibinfo{issn}{0733-8724}.

\bibitem[{\citenamefont{Radic and McKinstrie}(2003)}]{twopumpopt}
\bibinfo{author}{\bibfnamefont{S.}~\bibnamefont{Radic}} \bibnamefont{and}
  \bibinfo{author}{\bibfnamefont{C.~J.} \bibnamefont{McKinstrie}},
  \bibinfo{journal}{Optical Fiber Technology} \textbf{\bibinfo{volume}{9}},
  \bibinfo{pages}{7} (\bibinfo{year}{2003}).

\bibitem[{\citenamefont{Metcalfe
  et~al.}(2007{\natexlab{b}})\citenamefont{Metcalfe, Boaknin, Manucharyan,
  Vijay, Siddiqi, Rigetti, Frunzio, Schoelkopf, and Devoret}}]{CBA}
\bibinfo{author}{\bibfnamefont{M.}~\bibnamefont{Metcalfe}},
  \bibinfo{author}{\bibfnamefont{E.}~\bibnamefont{Boaknin}},
  \bibinfo{author}{\bibfnamefont{V.}~\bibnamefont{Manucharyan}},
  \bibinfo{author}{\bibfnamefont{R.}~\bibnamefont{Vijay}},
  \bibinfo{author}{\bibfnamefont{I.}~\bibnamefont{Siddiqi}},
  \bibinfo{author}{\bibfnamefont{C.}~\bibnamefont{Rigetti}},
  \bibinfo{author}{\bibfnamefont{L.}~\bibnamefont{Frunzio}},
  \bibinfo{author}{\bibfnamefont{R.~J.} \bibnamefont{Schoelkopf}},
  \bibnamefont{and} \bibinfo{author}{\bibfnamefont{M.~H.}
  \bibnamefont{Devoret}}, \bibinfo{journal}{Phys. Rev. B}
  \textbf{\bibinfo{volume}{76}}, \bibinfo{eid}{174516}
  (pages~\bibinfo{numpages}{5}) (\bibinfo{year}{2007}{\natexlab{b}}).

\bibitem[{\citenamefont{Yurke}(2004)}]{IOT}
\bibinfo{author}{\bibfnamefont{B.}~\bibnamefont{Yurke}}, in
  \emph{\bibinfo{booktitle}{Quantum Squeezing}}, edited by
  \bibinfo{editor}{\bibfnamefont{P.}~\bibnamefont{Drummond}} \bibnamefont{and}
  \bibinfo{editor}{\bibfnamefont{Z.}~\bibnamefont{Ficek}}
  (\bibinfo{publisher}{Springer}, \bibinfo{year}{2004}), pp.
  \bibinfo{pages}{53--95}.

\bibitem[{\citenamefont{Bergeal et~al.}(2008)\citenamefont{Bergeal, Vijay,
  Manucharyan, Siddiqi, Schoelkopf, Girvin, and Devoret}}]{bergeal-2008}
\bibinfo{author}{\bibfnamefont{N.}~\bibnamefont{Bergeal}},
  \bibinfo{author}{\bibfnamefont{R.}~\bibnamefont{Vijay}},
  \bibinfo{author}{\bibfnamefont{V.~E.} \bibnamefont{Manucharyan}},
  \bibinfo{author}{\bibfnamefont{I.}~\bibnamefont{Siddiqi}},
  \bibinfo{author}{\bibfnamefont{R.~J.} \bibnamefont{Schoelkopf}},
  \bibinfo{author}{\bibfnamefont{S.~M.} \bibnamefont{Girvin}},
  \bibnamefont{and} \bibinfo{author}{\bibfnamefont{M.~H.}
  \bibnamefont{Devoret}} (\bibinfo{year}{2008}), \eprint{arXiv.org:0805.3452}.

\bibitem[{\citenamefont{Clerk et~al.}(2003)\citenamefont{Clerk, Girvin, and
  Stone}}]{Clerk1}
\bibinfo{author}{\bibfnamefont{A.~A.} \bibnamefont{Clerk}},
  \bibinfo{author}{\bibfnamefont{S.~M.} \bibnamefont{Girvin}},
  \bibnamefont{and} \bibinfo{author}{\bibfnamefont{A.~D.} \bibnamefont{Stone}},
  \bibinfo{journal}{Phys. Rev. B} \textbf{\bibinfo{volume}{67}},
  \bibinfo{pages}{165324} (\bibinfo{year}{2003}).

\bibitem[{\citenamefont{Clerk et~al.}(2008)\citenamefont{Clerk, Devoret,
  Girvin, Marquardt, and Schoelkopf}}]{ClerkRMP2}
\bibinfo{author}{\bibfnamefont{A.~A.} \bibnamefont{Clerk}},
  \bibinfo{author}{\bibfnamefont{M.~H.} \bibnamefont{Devoret}},
  \bibinfo{author}{\bibfnamefont{S.~M.} \bibnamefont{Girvin}},
  \bibinfo{author}{\bibfnamefont{F.}~\bibnamefont{Marquardt}},
  \bibnamefont{and} \bibinfo{author}{\bibfnamefont{R.~J.}
  \bibnamefont{Schoelkopf}} (\bibinfo{year}{2008}),
  \eprint{arXiv.org:0810.4729}.

\bibitem[{\citenamefont{Dykman and Krivoglaz}(1994)}]{Dykman2}
\bibinfo{author}{\bibfnamefont{M.~I.} \bibnamefont{Dykman}} \bibnamefont{and}
  \bibinfo{author}{\bibfnamefont{M.~A.} \bibnamefont{Krivoglaz}},
  \bibinfo{journal}{Phys. Rev. E} \textbf{\bibinfo{volume}{49}},
  \bibinfo{pages}{1198} (\bibinfo{year}{1994}).

\bibitem[{\citenamefont{Walls}(1983)}]{squeezed1}
\bibinfo{author}{\bibfnamefont{D.~F.} \bibnamefont{Walls}},
  \bibinfo{journal}{Nature} \textbf{\bibinfo{volume}{306}},
  \bibinfo{pages}{141} (\bibinfo{year}{1983}).

\bibitem[{\citenamefont{Scully and Zubairy}(1997)}]{squeezed2}
\bibinfo{author}{\bibfnamefont{M.~O.} \bibnamefont{Scully}} \bibnamefont{and}
  \bibinfo{author}{\bibfnamefont{M.~S.} \bibnamefont{Zubairy}},
  \emph{\bibinfo{title}{{Quantum Optics}}} (\bibinfo{publisher}{Cambridge
  University Press}, \bibinfo{year}{1997}).

\bibitem[{\citenamefont{McKenzie et~al.}(2006)\citenamefont{McKenzie, Gray,
  Go\ss{}ler, Lam, and McClelland}}]{0264-9381-23-8-S31}
\bibinfo{author}{\bibfnamefont{K.}~\bibnamefont{McKenzie}},
  \bibinfo{author}{\bibfnamefont{M.~B.} \bibnamefont{Gray}},
  \bibinfo{author}{\bibfnamefont{S.}~\bibnamefont{Go\ss{}ler}},
  \bibinfo{author}{\bibfnamefont{P.~K.} \bibnamefont{Lam}}, \bibnamefont{and}
  \bibinfo{author}{\bibfnamefont{D.~E.} \bibnamefont{McClelland}},
  \bibinfo{journal}{Class. Quantum Grav.} \textbf{\bibinfo{volume}{23}},
  \bibinfo{pages}{S245} (\bibinfo{year}{2006}).

\bibitem[{\citenamefont{Braunstein and Kimble}(2000)}]{PhysRevA.61.042302}
\bibinfo{author}{\bibfnamefont{S.~L.} \bibnamefont{Braunstein}}
  \bibnamefont{and} \bibinfo{author}{\bibfnamefont{H.~J.}
  \bibnamefont{Kimble}}, \bibinfo{journal}{Phys. Rev. A}
  \textbf{\bibinfo{volume}{61}}, \bibinfo{pages}{042302}
  (\bibinfo{year}{2000}).

\bibitem[{\citenamefont{Slusher et~al.}(1986)\citenamefont{Slusher, Hollberg,
  Yurke, Mertz, and Valley}}]{PhysRevLett.56.788}
\bibinfo{author}{\bibfnamefont{R.~E.} \bibnamefont{Slusher}},
  \bibinfo{author}{\bibfnamefont{L.~W.} \bibnamefont{Hollberg}},
  \bibinfo{author}{\bibfnamefont{B.}~\bibnamefont{Yurke}},
  \bibinfo{author}{\bibfnamefont{J.~C.} \bibnamefont{Mertz}}, \bibnamefont{and}
  \bibinfo{author}{\bibfnamefont{J.~F.} \bibnamefont{Valley}},
  \bibinfo{journal}{Phys. Rev. Lett.} \textbf{\bibinfo{volume}{56}},
  \bibinfo{pages}{788} (\bibinfo{year}{1986}).

\bibitem[{\citenamefont{Yurke}(1987)}]{Yurke:87}
\bibinfo{author}{\bibfnamefont{B.}~\bibnamefont{Yurke}}, \bibinfo{journal}{J.
  Opt. Soc. Am. B} \textbf{\bibinfo{volume}{4}}, \bibinfo{pages}{1551}
  (\bibinfo{year}{1987}).

\bibitem[{\citenamefont{Yurke et~al.}(1988)\citenamefont{Yurke, Kaminsky,
  Miller, Whittaker, Smith, Silver, and Simon}}]{PhysRevLett.60.764}
\bibinfo{author}{\bibfnamefont{B.}~\bibnamefont{Yurke}},
  \bibinfo{author}{\bibfnamefont{P.~G.} \bibnamefont{Kaminsky}},
  \bibinfo{author}{\bibfnamefont{R.~E.} \bibnamefont{Miller}},
  \bibinfo{author}{\bibfnamefont{E.~A.} \bibnamefont{Whittaker}},
  \bibinfo{author}{\bibfnamefont{A.~D.} \bibnamefont{Smith}},
  \bibinfo{author}{\bibfnamefont{A.~H.} \bibnamefont{Silver}},
  \bibnamefont{and} \bibinfo{author}{\bibfnamefont{R.~W.} \bibnamefont{Simon}},
  \bibinfo{journal}{Phys. Rev. Lett.} \textbf{\bibinfo{volume}{60}},
  \bibinfo{pages}{764} (\bibinfo{year}{1988}).

\bibitem[{\citenamefont{Yurke et~al.}(1989)\citenamefont{Yurke, Corruccini,
  Kaminsky, Rupp, Smith, Silver, Simon, and Whittaker}}]{PhysRevA.39.2519}
\bibinfo{author}{\bibfnamefont{B.}~\bibnamefont{Yurke}},
  \bibinfo{author}{\bibfnamefont{L.~R.} \bibnamefont{Corruccini}},
  \bibinfo{author}{\bibfnamefont{P.~G.} \bibnamefont{Kaminsky}},
  \bibinfo{author}{\bibfnamefont{L.~W.} \bibnamefont{Rupp}},
  \bibinfo{author}{\bibfnamefont{A.~D.} \bibnamefont{Smith}},
  \bibinfo{author}{\bibfnamefont{A.~H.} \bibnamefont{Silver}},
  \bibinfo{author}{\bibfnamefont{R.~W.} \bibnamefont{Simon}}, \bibnamefont{and}
  \bibinfo{author}{\bibfnamefont{E.~A.} \bibnamefont{Whittaker}},
  \bibinfo{journal}{Phys. Rev. A} \textbf{\bibinfo{volume}{39}},
  \bibinfo{pages}{2519} (\bibinfo{year}{1989}).

\bibitem[{\citenamefont{Vijay}(2008)}]{Vijay}
\bibinfo{author}{\bibfnamefont{R.}~\bibnamefont{Vijay}}
  (\bibinfo{publisher}{Ph.D. Thesis, Yale University}, \bibinfo{year}{2008}).

\bibitem[{\citenamefont{Gerry and Knight}(2004)}]{bookGerry}
\bibinfo{author}{\bibfnamefont{C.~C.} \bibnamefont{Gerry}} \bibnamefont{and}
  \bibinfo{author}{\bibfnamefont{P.~L.} \bibnamefont{Knight}},
  \emph{\bibinfo{title}{Introductory Quantum Optics}}
  (\bibinfo{publisher}{{Cambridge University Press}}, \bibinfo{year}{2004}).

\bibitem[{\citenamefont{Koch et~al.}(2007)\citenamefont{Koch, Yu, Gambetta,
  Houck, Schuster, Majer, Blais, Devoret, Girvin, and Schoelkopf}}]{transmon}
\bibinfo{author}{\bibfnamefont{J.}~\bibnamefont{Koch}},
  \bibinfo{author}{\bibfnamefont{T.~M.} \bibnamefont{Yu}},
  \bibinfo{author}{\bibfnamefont{J.}~\bibnamefont{Gambetta}},
  \bibinfo{author}{\bibfnamefont{A.~A.} \bibnamefont{Houck}},
  \bibinfo{author}{\bibfnamefont{D.~I.} \bibnamefont{Schuster}},
  \bibinfo{author}{\bibfnamefont{J.}~\bibnamefont{Majer}},
  \bibinfo{author}{\bibfnamefont{A.}~\bibnamefont{Blais}},
  \bibinfo{author}{\bibfnamefont{M.~H.} \bibnamefont{Devoret}},
  \bibinfo{author}{\bibfnamefont{S.~M.} \bibnamefont{Girvin}},
  \bibnamefont{and} \bibinfo{author}{\bibfnamefont{R.~J.}
  \bibnamefont{Schoelkopf}}, \bibinfo{journal}{Phys. Rev. A}
  \textbf{\bibinfo{volume}{76}}, \bibinfo{pages}{042319}
  (\bibinfo{year}{2007}).

\bibitem[{\citenamefont{Dykman and Krivoglaz}(1984)}]{Dykman1}
\bibinfo{author}{\bibfnamefont{M.~I.} \bibnamefont{Dykman}} \bibnamefont{and}
  \bibinfo{author}{\bibfnamefont{M.~A.} \bibnamefont{Krivoglaz}},
  \bibinfo{journal}{Sov. Sci. Rev. A Phys.} \textbf{\bibinfo{volume}{5}},
  \bibinfo{pages}{265} (\bibinfo{year}{1984}).

\bibitem[{\citenamefont{Courty et~al.}(1999)\citenamefont{Courty, Grassia, and
  Reynaud}}]{Reynaud_qmfluc}
\bibinfo{author}{\bibfnamefont{J.-M.} \bibnamefont{Courty}},
  \bibinfo{author}{\bibfnamefont{F.}~\bibnamefont{Grassia}}, \bibnamefont{and}
  \bibinfo{author}{\bibfnamefont{S.}~\bibnamefont{Reynaud}},
  \bibinfo{journal}{Europhys. Lett.} \textbf{\bibinfo{volume}{46}},
  \bibinfo{pages}{31} (\bibinfo{year}{1999}).

\bibitem[{\citenamefont{Courty and Reynaud}(1992)}]{PhysRevA.46.2766}
\bibinfo{author}{\bibfnamefont{J.-M.} \bibnamefont{Courty}} \bibnamefont{and}
  \bibinfo{author}{\bibfnamefont{S.}~\bibnamefont{Reynaud}},
  \bibinfo{journal}{Phys. Rev. A} \textbf{\bibinfo{volume}{46}},
  \bibinfo{pages}{2766} (\bibinfo{year}{1992}).

\end{thebibliography}
\end{document}